\documentclass[11pt]{article}
\usepackage{epsf,amssymb,cite}
\def\paper{paper}

\newcommand{\xtra}[1]{{.}}
\renewcommand{\xtra}[1]{{, \tt hep-th/#1.}}
\newcommand{\xxtra}[1]{{, \tt hep-th/#1}}

\newcommand{\be}{\begin{equation}}
\newcommand{\ee}{\end{equation}}
\newcommand{\eq}{\begin{equation}}
\newcommand{\en}{\end{equation}}
\newcommand{\bea}{\begin{eqnarray}}
\newcommand{\eea}{\end{eqnarray}}

\newcommand\B[3]{{\ensuremath{ {}^{(#1)}\!B_{#2}^{#3}}}}
\newcommand{\ba}{\begin{array}}
\renewcommand{\bar}{\overline}
\newcommand{\blank}[1]{{}}
\newcommand{\mathematica}[1]{{}}
\newcommand{\bp}[3]{{\ensuremath{ \phi_{#1}^{(#2#3)}}}}

\newcommand\C[3]{{\ensuremath{ C_{#1 #2}{}^{#3} }}}

\newcommand{\CC}{{\hbox{\rm C\kern-0.5em{$\sf I$}}}}
\newcommand\cc[6]{{\ensuremath{ c^{(#1 #2 #3)}_{~~#4 #5}{}^{#6} }}}
\newcommand{\cev}[1]{\langle \,#1\,|}

\newcommand{\cg}{{\cal G}}
\newcommand{\cH}{{\cal H}}

\newcommand{\CPT}{{CPT}}

\newcommand{\ds}{\displaystyle}

\newcommand{\D}{{{\rm d}}}
\newcommand{\Dth}{{{\rm d}\theta}}

\newcommand{\dz}{{{\rm d}z}}

\newcommand{\e}{{\rm e}}
\newcommand{\ea}{\end{array}}

\newcommand{\ep}{\varepsilon}
\newcommand{\etaa}{}

\newcommand{\fract}[2]{{\textstyle\frac{#1}{#2}}}

\newcommand{\Ga}{\Gamma}

\newcommand{\gf}{$g$-function}
\newcommand{\cgf}{$\cg$-function}
\newcommand{\gfs}{$g$-functions}
\newcommand{\cgfs}{$\cg$-functions}

\renewcommand\hat{\widehat}

\newcommand{\icev}[1]{\langle\!\langle \,#1\,|}

\newcommand{\II}{\hbox{{\rm l{\hbox to 1.5pt{\hss\rm l}}}}}

\newcommand{\inti}{\int^{\infty}_{-\infty}}
\newcommand{\ivec}[1]{|\,#1\,\rangle\!\rangle}

\newcommand{\lf}{\left}

\newcommand{\m}{\phantom{-}}

\newcommand{\nn}{\nonumber}

\newcommand{\One}{{\hbox{{\rm 1{\hbox to 1.5pt{\hss\rm1}}}}}}
\renewcommand{\One}{{\mathbb 1}}
\renewcommand{\One}{{\rm 1\!\!1}}
\newcommand{\opnup}[1]{\renewcommand{\\}{\\[50 pt]}}

\newcommand{\resection}[1]{\setcounter{equation}{0}\section{#1}}
\newcommand{\ri}{\right}
\newcommand{\RR}{{\hbox{$\rm\textstyle I\kern-0.2em R$}}}

\newcommand{\te}{\theta}

\renewcommand{\tilde}{\widetilde}

\newcommand{\Tr}{{\rm Tr}}

\renewcommand{\vec}[1]{|\,#1\,\rangle}
\newcommand{\vev}[1]{\langle\,#1\,\rangle}
\newcommand{\veev}[2]{\langle\,#1\,|\,#2\,\rangle}
\newcommand{\Vev}[1]{\left\langle\,#1\,\right\rangle}
\newcommand{\Veev}[2]{\left\langle\,#1\,|\,#2\,\right\rangle}

\newcommand{\ZZ}{{\hbox{$\sf\textstyle Z\kern-0.4em Z$}}}

\newcommand{\CMP}[1]{Commun.\ Math.\ Phys.\ {\bf #1}}
\newcommand{\IJMP}[1]{Int.\ J.\ Mod.\ Phys.\ {\bf #1}}

\newcommand{\NP}[1]{Nucl.\ Phys.\ {\bf #1}}
\newcommand{\PL}[1]{Phys.\ Lett.\ {\bf #1}}

\newcommand{\PRL}[1]{Phys.\ Rev.\ Lett.\ {\bf #1}}

\newcommand{\ABZ}{A.B.~Zamolodchikov}
\newcommand{\AlBZ}{Al.B.~Zamolodchikov}

\begin{document}
%
\begin{titlepage}
\vskip 0.5cm
\begin{flushright}
DTP-99-67 \\
KCL-MTH-99-39 \\
ITFA 99-26 \\
{\tt hep-th/9909216}\\
September 29, 1999 \\
\end{flushright}
\vskip .8cm
\begin{center}
{\Large {\bf \gf\ flow in perturbed}} \\[5pt]
{\Large {\bf boundary conformal field theories } }
\end{center}
\vskip 0.8cm
\centerline{Patrick Dorey%
\footnote{e-mail: {\tt P.E.Dorey@durham.ac.uk}},
Ingo Runkel\footnote{e-mail: {\tt ingo@mth.kcl.ac.uk}},
Roberto Tateo\footnote{e-mail: {\tt tateo@wins.uva.nl}}
and G\'erard Watts\footnote{e-mail: {\tt gmtw@mth.kcl.ac.uk}}
}
\vskip 0.6cm
\centerline{${}^1$\sl Department of Mathematical Sciences,}
\centerline{\sl  University of Durham, Durham DH1 3LE,
England\,}
\vskip 0.2cm
\centerline{${}^3$\sl 
Universiteit van Amsterdam, Inst.~voor Theoretische Fysica\,}
\centerline{\sl
1018 XE Amsterdam, The Netherlands\,}
\vskip 0.2cm
\centerline{${}^{2,4}$\sl Mathematics Department, }
\centerline{\sl King's College London, Strand, London WC2R 2LS, U.K.}
\vskip 0.9cm
\begin{abstract}
\vskip0.15cm
\noindent
The \gf\ was introduced by Affleck and Ludwig  as a measure of the
ground state degeneracy of a conformal boundary condition.
We consider this function for perturbations of the conformal Yang--Lee
model by bulk and boundary fields using conformal perturbation
theory (\CPT), the truncated conformal space approach (TCSA) and
the thermodynamic Bethe Ansatz (TBA).
We find that the TBA equations derived by LeClair et al describe the
massless boundary flows, up to an overall constant, but are incorrect
when one considers a simultaneous bulk perturbation; however the TBA
equations do correctly give the  `non-universal' linear term in the
massive case, and the ratio of \gfs\ for different boundary conditions
is also correctly produced.
This ratio is related to the Y--system of the Yang--Lee model and
by comparing the perturbative expansions of the Y--system and of the
\gfs\ we obtain the exact relation between the UV and IR 
parameters of the massless perturbed boundary model.

\end{abstract}
\end{titlepage}
\setcounter{footnote}{0}
\def\thefootnote{\fnsymbol{footnote}}

\resection{Introduction}
Consider the partition function of a classical statistical-mechanical
system defined on a
\hbox{cylinder} of length $R$ and circumference
$L$. Among the characteristics of the model might be a bulk mass scale
$M$ and boundary scales depending on the boundary conditions $\alpha$
and $\beta$ imposed at the two ends of the cylinder; we will highlight
the role of these quantities by denoting the partition function
$Z_{\alpha\beta}(M,R,L)$.
If $R$ is taken to infinity with all other variables held fixed, then
\eq
  Z_{\alpha\beta}(M,R,L)
\sim
  A_{\alpha\beta}(M,L)\, \e^{ - R \,{E^{\rm circ}_0(M,L)}}
\;,
\label{lrasympt}
\en
where $E^{\rm circ}_0(M,L)$ is the ground state energy of the model
on a circle of circumference $L$. To derive this
asymptotic $R$-dependence, it is sufficient to treat the boundary
conditions as boundary states $\vec\alpha$ \cite{Card4} in a formalism
where time runs along the length of the cylinder, and states are
propagated by a bulk Hamiltonian $H_{\rm circ}(M,L)$:
\eq
  Z_{\alpha\beta}(M,R,L)
=   \cev {\alpha}
  \,\exp(-RH_{\rm circ}(M,L))\,
  \vec {\beta}
\;.
\label{lchan}
\en
At large $R$ the contribution of the ground state
$\vec\Omega$ dominates, establishing (\ref{lrasympt}) and
also giving
\eq
  A_{\alpha\beta}(M,L)
=   \frac
  {  \veev{\alpha}{\Omega}
  \, \veev{\Omega}{\beta}  }
  {\veev \Omega\Omega}
\;.
\label{eq:aml}
\en
The inner products appearing in (\ref{eq:aml})
should in general contain a  term
corresponding to a free-energy per unit length, i.e.
\eq
  \log(\, \frac{ \veev\Omega{\alpha} }{\veev\Omega\Omega^{1/2}}\, )
= -Lf_\alpha
+ \log(g_\alpha(M,L))
\;.
\label{logg}
\en
This linear term can in principle be extracted unambiguously from the
large $L$ behaviour of
$\log(\,  \veev \Omega {\alpha} \,)$.
The question is then whether the functions $\log(g_\alpha(M,L))$ now
contain universal information.

In the case $M= 0$, i.e.~for critical bulk, Affleck and
Ludwig~\cite{AL91} pointed out that the UV and IR limits of these
functions, 
$ \log( g_\alpha(0,0))$ and $\log( g_\alpha(0,\infty))$,
play the role of a generalised ground state degeneracy for
the UV and IR conformal boundary conditions respectively.
These can be easily calculated for many conformal field theories, and
can enable one to identify the boundary conditions uniquely.

The universality of the $g$-functions is however a somewhat delicate
issue when the model has a mass scale, either in the bulk or at the
boundary. For example, one could imagine that in
some calculational schemes the boundaries acquire a finite thickness,
and so the effective cylinder length 
would 
decrease by some finite amount $\delta$ to $R-\delta$, in which case
$\log(g_\alpha)$ would be altered 
\eq
  \log(g_{\alpha}(M,L))
\rightarrow
 \log(g_{\alpha}(M,L))+\frac{\pi\delta}{12L}c(M\!L)
\;,
\en
where $c(M\!L)$ is related to the ground state energy on the circle by
\eq
  E^{\rm circ}_0(M,L)
=  f_{\rm bulk}L-\frac{\pi}{6L}c(M\!L)
\;.
\en
More generally, the fact that the \gfs\ have to be extracted from
subleading contributions to $\log Z$ makes their calculation much
trickier than that of the corresponding bulk quantity $c(ML)$.  

If the model is integrable, one might hope to be able to address these
issues 
in
greater depth.
This was one of the topics of some work by
LeClair et al.\ \cite{LMSS}. Using the thermodynamic Bethe ansatz
(TBA) technique, they proposed equations for $\log(g)$, but up to now
these have not been checked in 
detail.
The main purpose of this \paper\ is to provide
some such checks in a concrete example, namely the
scaling Yang-Lee model,
making use both of conformal perturbation theory and of a
recently-introduced generalisation of the
TCSA to boundary situations~\cite{Us1}. 
We find that the TBA \gfs\ describes the {\em massless} flow well (up
to an overall constant) but the {\em massive} flows are not well
described.
However, we find that the TBA \gfs\ of \cite{LMSS} do appear to
describe correctly the {\em ratio} of the \gfs\ for different
boundary conditions of the same bulk model.
Further, the TBA expression for this ratio has a very simple form in
terms of the Y--function of the Yang--Lee model.
This enables us to (a) provide a perturbative expansion for the 
massive Y--function and (b) find the
coefficient in the relation between UV and IR parameters in the
boundary Yang-Lee model
model exactly, a result announced in \cite{Us1}.

We first discuss the conformal perturbation theory (\CPT) expressions
for $\log g$ and then how the truncated 
conformal space approach (TCSA)
can be used to investigate these functions.
We then turn to the 
TBA equations and the properties of their solutions.
Finally we compare the results of the two methods and find
the relation between UV and IR parameters.

\resection{Boundary conformal field theory}
\label{sec:bcft}

We begin with a sketch of boundary conformal field theory in general,
although we shall often specialise to those theories described in the bulk 
by a minimal Virasoro theories with `A'--type diagonal modular invariant.
At the end we give the details for the Yang-Lee model.
For a general discussion of the boundary content of the  Virasoro
minimal models see \cite{BPPZ}.

A boundary conformal field theory is specified by giving the field
content in the bulk, a list of the possible boundary conditions and
the field contents of these boundary conditions, the fields which
interpolate them, and some constants.

In the bulk there are primary fields $\Phi_i$ of weights $h_i,\bar
h_i$ which for a diagonal modular invariant are equal; under conformal
transformations they obey
\[
  \Phi_i(z,\bar z)
=  \left| \frac {\partial w}{\partial z} \right|^{x_i}
  \Phi_i(w,\bar w)
\;,
\]
where $x_i =  2 h_i$.

For the A--series, the conformal boundary conditions (which we shall
denote by $\alpha$) may themselves also be indexed by the Virasoro
minimal representations \cite{Card4,BPPZ}.
We adopt the notation $\bp j\alpha\beta (x)$
for a field of weight $h_j$
which interpolates two boundary conditions $\alpha,\beta$, which may
be the same or different. Such a field exists if the fusion
coefficient $N_{\alpha\,\beta}{}^i$ is one, and does not exist if it
is zero%
\footnote{Remember that for A-modular invariant Virasoro minimal
models all representations are self-conjugate and all the fusion
coefficients are either 1 or 0}.
We take such boundary primary fields to transform under conformal
transformations as%
\footnote{There is some choice in the transformations of boundary
fields, see \cite{RW}}%
\[
  \bp i \alpha \beta (z)
=  \left| \frac {\partial w}{\partial z} \right|^{h_i}
  \bp i \alpha \beta (w)
\;.
\]
Inside correlation functions we assume that the product of fields can
be replaced by their operator product expansion --  
we do not try to interpret them as genuine expansions of the actions
of operators, nor do we consider the associated  problems of orderings
of the arguments of the fields.

There are three different operator products -- bulk-bulk,
bulk-boundary and boundary-boundary.
The first  are the standard expressions%
\footnote{n.b. in the `A'-type invariant we only have spinless bulk
primary fields}  
,
\be
  \Phi_i(z,\bar z)\,
  \Phi_j(w,\bar w)
\sim
\sum_k \C ijk
       \left|z-w\right|^{x_k - x_i - x_j}
    \,  \Phi_k(w,\bar w)
+ \ldots
\;.
\label{ope1}
\ee
Next we have the operator products of the boundary primary fields:
\be
  \bp i \alpha\beta (z)\,
  \bp j \beta\gamma (w)
\sim
\sum_k \cc\alpha\beta\gamma ijk
       \,\left| z - w \right|^{h_k - h_i - h_j}
       \,\bp k \alpha\gamma (w) + \ldots
\;.
\label{ope2}
\ee
Finally, a primary bulk field near a boundary of type $\alpha$ may be
expanded in boundary fields as%
\be
  \Phi_i(z,\bar z)
\sim
\sum_j \B{\alpha}{i}{j}
       \, |2 (z - w)|^{h_j - x_i}
       \, \bp j \alpha\alpha (w)
  + \ldots
\;,
\label{ope3}
\ee
where $z$ is in the bulk and $w$ is the point on the boundary
closest to $z$.

To determine the correlation functions consistently, 
one needs to normalise the
different boundary conditions differently. One way to do this is to
give  the zero-point functions on a disc of radius~1 with boundary
condition $\alpha$, which  we denote by $Z_\alpha$,
\be
  Z_\alpha = {\Vev 1 }_\alpha^{\rm disc}
\;.
\label{z}
\ee
The rules for calculating the various constants in equations
(\ref{ope1}), (\ref{ope2}), (\ref{ope3}) and (\ref{z})
are essentially given in a series of papers by Cardy and Lewellen
\cite{Card4,CLew1,Lewe1}, but some work is required
to obtain explicit expressions.
For the structure constants of the
A--type model, see  \cite{Runk1}.

Using these structure constants, zero-point functions and the
chiral blocks, we can assign a value to any $n$-point function, i.e.\
to any surface with some assignment of conformal 
boundary conditions to the boundaries, and some insertions of bulk and
boundary fields; if the boundary conditions change at some point along
a boundary, then there is of necessity a boundary changing field
inserted at this point.

We can now express various simple correlation functions on a disc of
radius 1 in terms of these structure constants:
\be
\renewcommand{\arraystretch}{1.6}
 \begin{array}{rcl}
  \Vev {
  \bp i\alpha\alpha(e^{i\theta})
  }_\alpha^{\rm disc}
&\!\!=\!\!&
  0
\;,\;\;\;\;\;\;\;\;\;\;\;\;
  \Vev{\;1\;}_\alpha^{\rm disc} \;=\; Z_\alpha
\\[1mm] 
  \Vev {
  \Phi_i(  z , \bar z )
  }_\alpha^{\rm disc}
&\!\!=\!\!&
  \left(\,
     \B {\alpha}{i}{1}
  \, Z_\alpha
  \right)
  \, ( 1 - |z|^2 )^{- x_i}
\;,
\\[1mm] 
  \Vev {
  \Phi_i(  r e^{i\theta} )
  \;
  \bp j \alpha\alpha ( 1 )
  }_\alpha^{\rm disc}
&\!\!=\!\!&
  \left(\,
     \B {\alpha}{i}{j}
  \, \cc \alpha\alpha\alpha jj1
  \, Z_\alpha
  \right)
  \, \left|
    {1 {-} 2 r \cos\theta {+} r^2}
     \right|^{-h_j}
  \, \left|  1 {-} r^2 \right| ^{h_j - x_i}
\;,
\\[2mm] 
  \Vev {
  \bp i \alpha\beta ( {\theta_1})
  \;
  \bp i \beta\alpha ( {\theta_2}  )
  }^{\rm disc}
&\!\!=\!\!&
  \left(\,
  \cc \alpha\beta\alpha ii1
  \, Z_\alpha
  \right)
  \, \left| 2 \sin\!\fract{\theta_{12}}2  \right|^{-2 h_i}
\;,
\\[2mm] 
  \Vev {
     \bp i \alpha\beta  ( {\theta_1})
  \; \bp j \beta\gamma  ( {\theta_2})
  \; \bp k \gamma\alpha ( {\theta_3})
  }^{\rm disc}
&\!\!=\!\!&
  \left(\,
     \cc \alpha\beta\gamma ijk
  \, \cc \alpha\gamma\alpha kk1
  \, Z_\alpha
  \right)
\\ &\!\!{\times}\!\!&
   \left| 2 \sin\! \fract {\theta_{12} }2  \right|^{h_k - h_i - h_j}
   \left| 2 \sin\! \fract {\theta_{23} }2  \right|^{h_i - h_j - h_k}
   \left| 2 \sin\! \fract {\theta_{13} }2  \right|^{h_j - h_k - h_i}
\;,
\end{array}
\label{eq:nptfns}
\ee
where $\theta_{ij} =  \theta_i - \theta_j$, etc.

We are interested in calculating correlation functions on the cylinder
but we will be able to relate all the quantities of interest to
simpler correlation functions on a disc. 

Let us consider the partition function of a right cylinder of length
$R$, with the two ends being circles of circumference $L$ with
conformal boundary conditions $\alpha$ and $\beta$. This can be
realised as a rectangle in the upper half plane with vertices
$0$, $L$, $iR$ and $L + iR$, with the two vertical sides identified.
The partition function can be written as
\be
  Z_{(\alpha\beta)}^{\rm cyl}(R,L)
=  \Tr_{\cH_{(\alpha\beta)}}
   \left( \exp( - L \hat H^{\rm strip}_{(\alpha\beta)}(R) ) \right)
\;.
\ee

By mapping this rectangle to a half-annulus in the upper half plane by
$z \mapsto \exp( \pi z / R )$, we can use the operator formalism in
the upper-half plane. 
The Hilbert space $\cH_{(\alpha\beta)}$ decomposes into a direct
sum of irreducible representations of one copy of the Virasoro algebra
\be
  \cH_{(\alpha\beta)}
=  \oplus_i \, N_{(\alpha\beta)}{}^i\, \cH_i
\;,
\ee
and the Hamiltonian is given as 
\be
  \hat H_{(\alpha\beta)}^{\rm strip}
=  \frac \pi R ( L_0 - \frac c {24} )
\;,
\ee
so that
\be
  Z_{(\alpha\beta)}^{\rm cyl}(R,L)
=  \sum_i \, N_{\alpha\beta}{}^i\, \chi_i(q)
\;,
\label{eq:z1}
\ee
where $\chi_i(q)$ is the character of the representation $i$,
$q =  \exp( - \pi L / R)$, and $N_{\alpha\beta}{}^i$ are the
Verlinde fusion numbers.

We can also map the cylinder into a complete annulus 
of inner and outer radii $\exp(-2\pi R/L)$ and $1$ respectively
by $z {\mapsto} w {=} \exp( 2 i  \pi z / L )$,
and use the operator formalism on the plane.
The Hilbert space for the bulk theory on the plane
carries a representation of two copies of the Virasoro algebra, in
terms of which the cylinder Hamiltonian is given as
\be
  \hat H^{\rm circ}
= \frac{2 \pi}{L} ( L_0 + \bar L_0 - \frac c{12} )
\;.
\label{eq:cylham}
\ee
This space carries a  
representation of two copies of the Virasoro algebra.
The highest weight states $\vec i$ are labelled by the highest weight
representations of the Virasoro algebra and given by the action of the
bulk primary fields on the $SL(2)$--invariant vacuum $\vec 0$ by
\be
  \vec i
=  \lim_{w \to 0}
  \Phi_i(w,\bar w) \, \vec 0
\;,
\label{eq:veci}
\ee
and hence \[
  \veev i j
=  \delta_{ij}
  \, \C ii1
  \,\veev 0 0
\;.
\]
We shall not necessarily normalise the vacuum to be norm 1, and indeed
for the Yang-Lee model it is convenient to take
$\veev 0 0 =  -1$.

The partition function on the cylinder can be calculated 
in terms of boundary states $\vec \alpha$, $\vec \beta$
in the Hilbert space for the bulk theory on the plane:
\be
  Z^{\rm cyl}_{\alpha\beta}
= \cev\alpha \, e^{-R \hat H^{\rm circ}(L)} \, \vec\beta
\;.
\label{eq:cylpart}
\ee
For a boundary state to represent a circular boundary
with conformally invariant boundary condition, it must satisfy
\[
  \left(  L_m - \bar L_{-m} \right)
  \vec{  \alpha }
=  0
\;,\;\;
  \cev{  \alpha }\left( L_m - \bar L_{-m} \right)
=  0
\;.
\]
A basis of states satisfying these conditions are the
Ishibashi states. There is one such state for each Virasoro
representation in the minimal model in question, which we denote by 
\[
  \ivec{ i}
= \left\{
  1 + \frac{ L_{-1} \bar L_{-1} }{2 h_i} + \ldots
  \right\}
  | i \rangle
\;,\;\;\;\;
  \ivec{ 0 }
= \left\{
  1 + \frac{ L_{-2} \bar L_{-2} }{c/2} + \ldots
  \right\}
  | 0 \rangle
\;.
\]
\[
  \icev{ i}
= \cev i 
  \left\{
  1 + \frac{ L_{1} \bar L_{1} }{2 h_i} + \ldots
  \right\}
\;,\;\;\;\;
  \icev{ 0 }
= \cev 0
  \left\{
  1 + \frac{ L_{2} \bar L_{2} }{c/2} + \ldots
  \right\}
\;.
\]
These are normalised so that 
\[
  \icev i \,
  e^{ - R \hat H^{\rm circ}(L) }
  \, \ivec j
= \delta_{ij}\, \chi_i(\tilde q)\, \veev i i 
\;,
\]
where $ \tilde q  = \exp( - {4 \pi R}/ L ) $.
We expand the boundary states in terms of the Ishibashi states as
\[
  \vec {\alpha}
=  \sum_j g_\alpha^j
  \,\ivec {j}
\;.
\]
(It is always possible to normalise the fields and states such that
the coefficients $g_\alpha^i$ are real, and we shall assume that has
been done. See \cite{Runk1} for details).
Using these states, the partition function 
is given from (\ref{eq:cylpart})  by
\bea
  Z_{(\alpha\beta)}^{\rm cyl}(r_1,r_2)
&= & \sum_i
  g_\alpha^i\, g_\beta^i\, 
  \chi_i(\tilde q)\,
  \veev i i 
\;.
\label{eq:z2}
\eea
Noting that 
\[
  Z_{\alpha\beta}^{\rm cyl}
= \sum_{ij}
  N_{\alpha\beta}^i\, S_i{}^j\, \chi_j(\tilde q)
\;,
\]
where $S_i{}^j$ is the matrix implementing the modular transformation,
we arrive at Cardy's result,
\be
  g_\alpha^i\, g_\beta^i\, \veev i i
=  \sum_j
  N_{\alpha\beta}{}^j \,
  S_{j}{}^{i}
\;.
\label{eq:ggs}
\ee
These equations can then be solved to find the $g_\alpha^i$.

One can also show that the same boundary states $\vec\alpha$ describe
a circular boundary of radius one in the plane%
\footnote{n.b. the normalisations of annulus partition functions
differ from those of cylinder partition functions 
\cite{CPes1}, so some care must be taken to define boundary states for
discs of general radius.
}
and so the constants $Z_\alpha$ can now be found
by expressing the one-point functions on a disc of radius 1 in two
different ways as 
\be
  \B \alpha i \One \,Z_\alpha
= \Vev{\Phi_i(0)}_\alpha^{\rm disc}
= \veev {\alpha} i
= \sum_j g_\alpha^j\, \icev {j} \, i\, \rangle
=  g_\alpha^i\, \veev ii
\;,
\ee
to find 
\be
  Z_\alpha   \, 
  \B\alpha i\One 
= g_\alpha^i \,
  \veev ii 
\;.
\label{eq:zb=gii}
\ee
This completes the identification of the structure constants and
zero-point functions. To end this section we recall how to find
the conformal values of the \gfs\ in terms of this conformal data.
To fix the \gfs\ $g_\alpha$, we note that the leading term in the
cylinder partition function is given by
\be
  Z_{(\alpha\beta)}^{\rm cyl}(R,L)
\sim
  g_\alpha^\Omega\,
  g_\beta^\Omega\,
  e^{ - R E_{\Omega} }\,
  \veev \Omega\Omega\,
\;.
\ee
where $\vec\Omega$ is the state of lowest conformal weight in the
theory defined on the plane.
This means that the \gfs\ $g_\alpha$ satisfy
\be
  g_\alpha\,g_\beta
= g_\alpha^\Omega\,
  g_\beta^\Omega\,
  \veev \Omega\Omega
\;.
\label{eq:gab}
\ee
However, recalling that the conformal vacuum representation `$0$'
satisfies 
$  N_{0\alpha}{}^i = \delta_\alpha^i $,
we see from (\ref{eq:ggs}) that
\be
  (g_\One^\Omega)^2 
= \frac {S_{\Omega 0}}{\veev \Omega\Omega}
\;,\;\;\;\;
   g_\One^\Omega\,g_\alpha^\Omega 
= \frac {S_{\Omega\alpha}}{\veev \Omega\Omega}
\;.
\label{eq:goO}
\ee
Finally, we recall that the modular $S$-matrix for the Virasoro
minimal models  satisfies $S_{\Omega\alpha} {>} 0$ for all
representations $\alpha$, and so if we require the \gfs\ to be
positive and the $g_\alpha^i$ to be real we should take 
$\veev \Omega\Omega >0$ and can read off their 
values from eqns.\ (\ref{eq:gab}) and (\ref{eq:goO}) as
\be
  g_\One
= | S_{\Omega 0} |^{1/2}
\;,\;\;\;\;
  g_\alpha 
= \frac {S_{\Omega\alpha}~~~}
        { | S_{\Omega 0} |^{1/2}}
\;,\;\;\;\;
  g_\alpha^\Omega 
= \frac{g_\alpha }{\sqrt{ \veev \Omega \Omega~ }}
\;.
\ee
We now give the explicit results for the case of the Yang-Lee model.

\subsection{The Yang-Lee model}

The Yang-Lee model is the simplest non-unitary conformal field theory,
$M_{2,5}$, and has central charge $-22/5$ and effective central charge 
$2/5$. There are only two representations of the Virasoro algebra of
interest, of weight $0$ and $-1/5$, and consequently only two bulk
primary fields, the identity $\One$ of weight $0$, and $\varphi$ of
weight $x_\varphi =  -2/5$; equally there are only two
conformally-invariant boundary conditions, which we denote by $\One$
and $\Phi$. The fusion rules are
\be
\begin{array}{rcl}
  \One \times \One &= & \One
\;,
\\
  \One \times \varphi &= & \varphi
\;,
\\
  \varphi \times \varphi &= & \One + \varphi
\;.
\end{array}
\label{eq:fr}
\ee
{}From these rules we read off that there are only three non-trivial
boundary fields, all of weight $h_\phi = 1/5$. Two of these
interpolate the two different conformal boundary conditions,
\be
  \bp {-1/5}\Phi\One   
\;,\;\;\;\;
  \bp {-1/5}\One\Phi
\;,
\ee
which we shall both denote by $\psi$,
and one lives on the $\Phi$ boundary,
\be
  \phi
\equiv
  \bp {-1/5}\Phi\Phi
\;.
\ee
As indicated%
\footnote{
In this model, there is only a single non-trivial bulk field, a single 
boundary condition admitting a relevant perturbation, and a single
non-trivial boundary field, all labelled by the same Virasoro
representation, and hence we have denoted them by $\varphi, \Phi$ and
$\phi$ respectively, in the hope that this will reduce confusion.}%
, we shall often simply denote the non-trivial
boundary primary field on the $\Phi$ boundary by $\phi$.

If we label the two representations of the Virasoro algebra by $\phi$
and 1 then the modular $S$-matrix is
\be
  S
= \pmatrix{ S_{11} & S_{1\phi} \cr S_{\phi 1} & S_{\phi\phi}}
= \fract 2{\sqrt 5}\,
  \pmatrix{ -\sin\fract{2\pi}5 &  \sin\fract\pi 5 \cr
            ~\sin\fract\pi 5   & \vphantom{\Big|}~\sin\fract{2\pi}5 }
= \pmatrix{ -0.8506.. & 0.5257.. \cr ~~0.5257.. & 0.8506.. }
\;.
\ee

\subsubsection{Yang--Lee structure constants}
\label{sec:ylscs}

We give here the structure constants  appearing in all the
operator products of interest, that is the bulk OPE 
\bea
  \varphi(z,\bar z)\;
  \varphi(w,\bar w)
&=&
  \C\varphi\varphi\One \, |z-w|^{4/5}
\;+\; \C\varphi\varphi\varphi \, |z-w|^{2/5}\,\varphi(w,\bar w)
\;+\; \ldots
\;,
\nn\\
\noalign{%
\vskip 1mm%
\noindent the boundary OPEs,%
\vskip 1mm}
  \phi(z)\;\phi(w)
&=&
  \C\phi\phi\One \, |z-w|^{2/5}
\;+\; \C\phi\phi\phi \, |z-w|^{1/5}\,\phi(w)
\;+\; \ldots
\;,
\nn\\
  \phi(z)\;\psi(w)
&=&
  \C\phi\psi\psi
  |z-w|^{1/5}\,
  \psi(w)
\;+\; \ldots
\;,
\nn\\
\noalign{%
\vskip 1mm%
\noindent and the two bulk--boundary OPEs%
\vskip 2mm}
 \left. \varphi(z)\;\right|_\One
&=&
  \B\One\varphi\One \, |2(z-w)|^{2/5}
\;+\;\ldots
\;,
\nn\\
 \left. \varphi(z)\;\right|_\Phi
&=&
  \B\Phi\varphi\One \, |2(z-w)|^{2/5}
\;+\;
 \B\Phi\varphi\phi \, |2(z-w)|^{1/5}\,\phi(w)
\;+\; \ldots
\;.
\nn
\eea
We want all these structure constants to be real, and a suitable
choice is
\be
{
\renewcommand{\arraystretch}{1.7}
\begin{array}{rclrcl}
\multicolumn{6}{c}{
\C\varphi\varphi\One
 ~=~ \C\phi\phi\One
 ~=~
  -1
\;,
}

\\

  \C\varphi\varphi\varphi
&= &
  - \left|
    \fract{2}{1 + \sqrt 5}
    \right|^{1/2}
    \cdot\alpha^2
\;,
&

  \B \One\varphi\One
&= &
  -\left| \fract{2}{1 + \sqrt 5} \right|^{1/2}
\;,
\\

  \C\phi\phi\phi
&= &
  - \left|
    \fract{1 + \sqrt 5}{2}
    \right|^{1/2}
    \cdot\alpha
\;,
&

  \B \Phi\varphi\One
&= &\m
  \left| \fract{1 + \sqrt 5}{2} \right|^{3/2}
\;,
\\

  \C\phi\psi\psi
&= &
  - \left|
    \fract{2}{1 + \sqrt 5}
    \right|^{1/2}
    \cdot\alpha
\;,
&

  \B \Phi\varphi\phi
&= &\m
  \left|
    \fract{5 + \sqrt 5}2
  \right|^{1/2}
  \cdot\alpha
\;,
\\

\multicolumn{6}{c}{
  \alpha
 ~=~
  \m
  \left|
  \fract{\Ga(1/5)\,\Ga(6/5)}{\Ga(3/5)\,\Ga(4/5)}
  \right|^{1/2}
\;.
}

\end{array}
}
\label{eq:ylscs}
\ee
The state of lowest conformal dimension in the bulk theory is
\be
  \vec\Omega = \vec\varphi
\;,
\ee
and we choose to normalise the bulk highest-weight states as
\be
  \veev 0 0 = -1
\;\;\;\;\Rightarrow\;\;\;\;
  \veev\varphi\varphi = 1
\;.
\ee
With this, and demanding $g_\alpha^\varphi>0$, the coefficients appearing
in the boundary states are
\be
\begin{array}{rcrcrcr}
  g_\One^0       &=&  \left| \fract{\sqrt 5 +1}{2\sqrt 5} \right|^{1/4}
&\;,\;\;&
  g_\One^\varphi &=&  \left| \fract{\sqrt 5 -1}{2\sqrt 5} \right|^{1/4}
\\
  g_\Phi^0       &=& -\left| \fract{\sqrt 5 -2}{\sqrt 5} \right|^{1/4}
&\;,\;\;&
  g_\Phi^\varphi &=&  \left| \fract{2 + \sqrt 5}{\sqrt 5} \right|^{1/4}
\end{array}
\ee
The zero-point functions are then given as
\be
  Z_\One
=  {\Vev 1}_\One^{\rm disc}
= \veev{\One} 0
= -g_\One^0
\;,\;\;\;\; 
  Z_\Phi
=  {\Vev 1}_\Phi^{\rm disc}
= \veev{\Phi} 0
= -g_\Phi^0
\;.
\ee
Finally, the \gfs\ are given by 
$
  g_\alpha
= g_\alpha^\Phi \sqrt{\Veev \varphi \varphi}
= g_\alpha^\Phi
$, i.e.
\be
\begin{array}{rlrl}
  \log g_\One
&
= \;
\fract 14 \log\left| \fract{ \sqrt 5 - 1 }{2 \sqrt 5} \right|
\; =
&\!\! - \fract 14 \log\left| \fract{1 + \sqrt 5}2 \right|
  - \fract 18 \log 5
&
= -0.321482..
\;,
\\[1mm]
  \log g_\Phi
&
= \;
\fract 14 \log\left| \fract{ 2 + \sqrt 5 }{\sqrt 5} \right|
\; =
&\!\! \fract 34 \log\left| \fract{1 + \sqrt 5}2 \right|
  - \fract 18 \log 5
&
= \m 0.159729..
\;.
\end{array}
\label{eq:gvalues}
\ee

\subsubsection{Yang--Lee correlation functions}
\label{sec:ylcfs}

In terms of the structure constants
(\ref{eq:ylscs}), the simple correlation functions 
(\ref{eq:nptfns}) on a disc of unit radius are
\be
{
\renewcommand{\arraystretch}{1.4}
\begin{array}{rcl}

    \Vev{\;1\;}^{\rm disc}_\alpha
&=& Z_\alpha
\;,\;\;\;\;\;\;\;\;

    \Vev{\;\phi(e^{i\theta})\;}^{\rm disc}_\Phi
\;=\; 0
\\

    \Vev{\;\varphi(\, re^{i\theta})\;}^{\rm disc}_\alpha
&=& \B\alpha\varphi\One\, 
    Z_\alpha\,
    (1 - r^2)^{2/5}
\\

    \Vev{\; \phi(e^{i\theta})\;\phi(1)\;}^{\rm disc}_\Phi
&=& \C\phi\phi\One\,
    Z_\Phi\,
    ( 2 \sin(\theta/2) )^{2/5}
\\

    \Vev{\; \varphi(\,r e^{i\theta}\,)
         \; \phi(1) \;}^{\rm disc}_{\Phi}
&=& \B\Phi\varphi\phi\,
    \C\phi\phi\One\,
    Z_\Phi\,
    ( 1 - 2 r \cos\theta + r^2)^{1/5}
    ( 1 - r^2)^{1/5}

\end{array}
}
\label{eq:ylcfs}
\ee
We shall also need two more correlation functions, which are given
in terms of chiral four-point blocks.
A basis for the chiral four-point functions of the weight
$-1/5$ field are $f^1$ and $f^\phi$, 
given in terms of hypergeometric functions by
\be
  f^1(x) 
= [x(1-x)]^{2/5} \,
  F(\fract 35,\fract 45;\fract 65;x)
\;,\;\;\;\; 
  f^\phi(x) 
= x^{1/5}\,
 (1-x)^{2/5} \,
  F(\fract 25,\fract 35;\fract 45;x)
\;.
\ee
In terms of these chiral blocks we can give the 
bulk two-point functions on  a disc of radius 1:
\bea
    \Vev{   \varphi(0)\,\varphi(re^{i\theta})   }_\alpha^{\rm disc}
&=& Z_\alpha\,\left\{\;
    \C\varphi\varphi\One   f^1(r^2)
  \;+\;
    \B\alpha\varphi\One\,\C\varphi\varphi\varphi  f^\phi(r^2)
  \; \right\}
\;,
\\
\noalign{\vskip 1mm%
\noindent
which is rotationally invariant.
We can also give the 1-bulk--2-boundary point functions in terms of
these two chiral blocks, but it is more convenient to perform some
transformations on the arguments to make the expressions manifestly
real, in which case we find
\vskip 2mm}
   \Vev{ \varphi(0)\; \phi(1)\; \phi(e^{i\theta}) }_\Phi^{\rm disc}
 \!\!\!\!
 &\!\!{=}\!\!&\!\!
  Z_\Phi \, \C \phi\phi\One\,
  \Big\{\;
  \B\Phi\varphi\One\,
   (2\sin\fract\theta2)^{2/5}\,
   (\cos\fract\theta2)^{-4/5}\,
   F(\fract{4}{10},\fract{9}{10};\fract{11}{10};\,{-}\tan^2\!\fract\theta 2)  
\nn\\ 
 &\!\!{+}\!\!&\!\!
   \B\Phi\varphi\phi\,\C \phi\phi\phi
   (2\sin\fract\theta2)^{1/5}\,
   (\cos\fract\theta2)^{-3/5}\,
   F(\fract{3}{10},\fract{8}{10};\fract{9}{10};\,{-}\tan^2\!\fract\theta 2)  
\; \Big\}
   \;.
\nn\\
\eea

\resection{%
\gfs\ and perturbed conformal field theory
}
\label{sec:gpcft}

We define the \cgfs\ of the perturbed theory as
\be
  \cg_\alpha
= \frac{ \veev \alpha \Omega }{ \veev \Omega\Omega^{1/2}}
\;,
\label{eq:cgdef}
\ee
where $\vec\alpha$ is the (possibly perturbed) boundary state and
$\vec\Omega$ is the ground state of the perturbed theory.
Typically these functions will not be equal to the \gfs\
$g_\alpha$ but will differ by the extensive free--energy
terms discussed in the introduction:
\[
  \log \cg_\alpha 
= \log g_\alpha \; - \; L \, f^{\rm pcft}_\alpha
\;.
\]
However it is the $\cg_\alpha$ that are accessible in conformal perturbation
theory, and not the $g_\alpha$. Currently it is not possible to determine
$f^{\rm pcft}_\alpha$ directly from conformal perturbation theory, but
using the truncated conformal space approximation (TCSA), it is
possible to approximate the functions $\cg_\alpha$ numerically for a
reasonable range of $L$, and this will enable us in some cases to
estimate $f^{\rm pcft}_\alpha$.
For the case of the Yang--Lee model, 
the boundary conditions of
interest are $\One$ (which admits no relevant perturbation) and
$\Phi(h)$, the perturbation of the conformal $\Phi$ boundary by the
integral along the boundary  
\be
  h \int \phi(x) \; \D x
\;.
\label{eq:boundpert}
\ee
The bulk perturbation corresponds to a term in the action
\be
  \lambda \int \varphi(w,\bar w) \; \D^2 w
\;.
\label{eq:bulkpert}
\ee
The functions $\cg_\alpha$ therefore have expansions
\be
{\renewcommand{\arraystretch}{1.4}
\begin{array}{rcl}
    \log \cg_\One(\lambda,L)
&=& \sum_{n=0}^\infty d_n
 \, (\lambda L^{12/5})^n
\;,
\\
    \log \cg_{\Phi(h)}(\lambda,L)
&=& \sum_{m,n=0}^\infty c_{mn} 
 \, ( h L^{6/5} )^m 
    (\lambda L^{12/5})^n
\;.
\end{array}
}
\label{eq:cmndn}
\ee
In section \ref{sec:btcsa} we discuss the TCSA calculation of
$\cg_\alpha$, but before that, 
in the next section we calculate the following 
coefficients in their expansions:
\be
{\renewcommand{\arraystretch}{1.4}
\begin{array}{rcl}
    \log \cg_\One(\lambda,L)
&=& \log g_\One 
 \;+\;  d_1 \, (\lambda L^{12/5})
 \;+\; \ldots
\;,
\\
    \log \cg_{\Phi(h)}(\lambda,L)
&=& \log g_\Phi 
 \;+\;  c_{10}\, (h L^{6/5})
 \;+\;  c_{20}\, (h L^{6/5})^2
 \;+\;  c_{01}\, (\lambda L^{12/5})
 \;+\; \ldots
\end{array}
}
\label{eq:cddefs}
\ee
To the order to which we will work, the bulk and boundary
perturbations can be treated independently, so we shall first
calculate $c_{01}$ and $d_1,$ and then $c_{10}$ and $c_{20}$. 

\subsection{%
The bulk perturbation
}
\label{sec:bulkp}

For a purely bulk perturbation (\ref{eq:bulkpert}), 
after mapping to the plane, the Hamiltonian of the perturbed theory is 
\bea
  \hat H^{\rm circ}(\lambda, L)
&=&
  \hat H_0 
\;+\;
  \lambda \, \hat H_1
\nn\\
&=&
  \frac{2 \pi}{L} ( L_0 + \bar L_0 - \frac c{12})
\;+\;
  \lambda \,
  \left( \frac{L}{2\pi} \right)^{7/5} 
  \oint
  \varphi(e^{i\theta} )\, \D\te
\;.
\label{eq:cylham2}
\eea
The full ground state is given by
\be
  \vec\Omega
= \vec\varphi
+ {\lambda \sum_a}' \Omega_a \vec{\psi_a}
+ O(\lambda^2)
\;,
\label{eq:fgs}
\ee
where the sum is over an orthonormal eigenbasis of $\hat H_0$
(where the $'$ indicates that the ground state $\vec\varphi$ is to be
excluded) and the coefficients $\Omega_a$ are given by first order
perturbation theory as 
\be
  \Omega_a
= \frac{\cev{\psi_a} \, \hat H_1 \, \vec\varphi}
       {\cev\varphi \hat H_0 \vec\varphi - \cev{\psi_a} \hat H_0 \vec{\psi_a}}
\;.
\label{eq:omi}
\ee
Combining (\ref{eq:omi}) and (\ref{eq:cylham2}), we find
\be
  {\sum_a}' \Omega_a \vec{\psi_a}
= - 
  \left( \frac L{2\pi} \right)^{12/5} 
  \, (1{-}P)
  \,\frac{1}{ L_0 + \bar L_0 + 2/5 }
  \, (1{-}P)
  \oint
  \varphi(e^{i\theta} )\, 
  \vec\varphi
  \, \D\te
\;,
\label{eq:om2}
\ee
where $P = \vec\varphi\cev\varphi$ is the projector onto the state
$\vec\varphi$. 
{}From (\ref{eq:fgs}), the norm of $\vec\Omega$ is unchanged to order
$\lambda$, and hence
the correction to the \cgf\ is given by
\be
  \log \frac{ \cg_\alpha(\lambda,L) }{ \cg_\alpha(0,L) }
= \lambda L^{12/5} a_\alpha + O(\lambda^2)  
\;,
\label{eq:adef}
\ee
where
\be
  a_\alpha
= - \frac{1}{g_\alpha}
  (2\pi)^{-12/5}
  \oint \D\te 
  \cev\alpha
  (1{-}P)
  \frac{1}{ L_0 + \bar L_0 + 2/5 }
  (1{-}P)
  \,\varphi(e^{i\te})
  \,\vec\varphi
\;.
\label{eq:aa1}
\ee
Using the rotational invariance of the boundary state and of
$\cev\varphi$ and the relations
\[
  \int_0^1 \!x^{\alpha}\, \D x = \frac{1}{\alpha+1}
\;,\;\;\;\;
   x^{L_0 + \bar L_0} \, 
   \varphi(1)
   \, x^{- L_0 - \bar L_0} \, 
=  x^{- 2/5} \, \varphi(x) 
\;,\;\;\;\;
  \cev\alpha  P
= g_\Phi \cev\varphi
\;,
\]
we arrive at
\be
  a_\alpha
= - (2\pi)^{-7/5}
  \int_0^1 \frac{\D x}{x^{7/5}}
  \left(   
  \frac{1}{g_\alpha} \cev\alpha \varphi(x) \vec\varphi
  \;-\;
  \cev\varphi \varphi(x) \vec\varphi
  \right)
\;.
\label{eq:aa2}
\ee
The correlation functions in (\ref{eq:aa2}) are 
\bea
  \cev\alpha\, \varphi(x)\, \vec\varphi
&=& \Vev{ \varphi(x) \varphi(0) }^{\rm disc}_{\alpha}
\;=\;
  Z_\alpha\,\left\{\;
    \C\varphi\varphi\One   f^1(x^2)
  \;+\;
    \B\alpha\varphi\One\,\C\varphi\varphi\varphi  f^\phi(x^2)
  \; \right\}
\nn\\
  \cev\varphi\, \varphi(x)\, \vec\varphi
&=& - \C\varphi\varphi\varphi
  \, x^{2/5}
\;,
\nn
\eea
so that using (\ref{eq:zb=gii}), 
the expression for $a_\alpha$ simplifies to
\be
  a_\alpha 
= - (2\pi)^{-7/5}
  \Big(
  \C\varphi\varphi\varphi\, I_2
\;\;+\;\;
  \frac{\C\varphi\varphi\One}{\B\alpha\varphi\One}\, I_1
  \Big)
\;,
\ee
where the integrals $I_1$ and $I_2$ are
given (by Mathematica) as 
\bea
I_1 &=&
  \int_0^1 \frac{{\rm d}x }{x^{7/5}}
  f^1( x^2)
\;=\;
  \int_0^1 {\rm d} x \;
   x^{-3/5}
  (1 -  x^2)^{2/5}
  F(\fract 35,\fract 45;\fract 65;  x^2)
\nn\\
&=& 
    \frac {\pi}{10 \sin(2\pi/5)}
    \frac { \Gamma(\fract 15)\, \Gamma(\frac 25)}
          { \Gamma(\fract 45)^2 }
\;=\;
   2.48171...
\mathematica{   2.481712549737395
 N [ 
 (Pi / ( 10 Sin[ 2 Pi/5])) * 
 Gamma[1/5] Gamma[2/5] / Gamma[4/5]^2
, 16 ]
            }
\;,
\nn\\
\noalign{\vskip 1mm%
\noindent%
and the somewhat obscure expression
\vskip 2mm
}
 I_2 
&=&
  \int_{0}^1\! \frac{{\rm d} x }{ x^{7/5}}
      (f^\phi( x^2) {-}  x^{2/5})
\; = \;
\mathematica{ 
  Integrate[
  (1/r)(-1 + (1-r^2)^(2/5) Hypergeometric2F1[2/5,3/5,4/5,r^2]),
  {r,0,1}]
}
   \int_0^1 \frac{{\rm d} x } x 
   \left(
   (1- x^2 )^{2/5}F(\fract 25,\fract 35; \fract 45;  x^2 ) 
   -   1
   \right)
\nn\\
&=&
\mathematica{
 (5/56) (-14 - 3*HypergeometricPFQ[{1, 7/5, 8/5}, {9/5, 12/5}, 1] ) + 
 (Pi (5 + Sqrt[5])^(3/2) )/( 8 Sqrt[10] ) +
 (5/8) Log[5] - Sqrt[5] Log[ (1 + Sqrt[5])/2]/4
(**)
 - 5/4 - (15/56)*HypergeometricPFQ[{1, 7/5, 8/5}, {9/5, 12/5}, 1]  + 
 (Pi (5 + Sqrt[5])^(3/2) )/( 8 Sqrt[10] ) +
 (5/8) Log[5] - Sqrt[5] Log[ (1 + Sqrt[5])/2]/4
}
 - \fract{15}{56}\, {}_3F_2(1,\fract 75,\fract85;\fract{9}5,\fract{12}5 ;1) 
 + \fract{\pi}{8}\,\fract{~( 5 + \sqrt 5 )^{3/2}\hskip -3mm }{\sqrt{10~}}
\;- \fract{\sqrt 5}{~4} \log\!\fract{1 + \sqrt 5}{2}
 + \fract 58 (\log 5 - 2)
\nn\\
&=& -0.083937990...
\;.
\eea
Substituting these values into $a_\alpha$ we find
\bea
  d_1
\;=\;
   a_\One
&{\!\!=\!\!}&
   - (2\pi)^{-7/5}
  \Big(
  \C\varphi\varphi\varphi\, I_2
\;\;+\;\;
  \frac{\C\varphi\varphi\One}{\B\One\varphi\One}\, I_1
  \Big)
\;=\;
  -0.25311758..
\;,
\label{eq:d1}
\\
  c_{01}
\;=\;
   a_\Phi
&{\!\!=\!\!}&
   - (2\pi)^{-7/5}
  \Big(
  \C\varphi\varphi\varphi\, I_2
\;\;+\;\;
  \frac{\C\varphi\varphi\One}{\B\Phi\varphi\One}\, I_1
  \Big)
\;=\;
  \m 0.0797648257..
\;.
\label{eq:c01}
\eea
While the expression for $I_2$ is rather unwieldy, it cancels exactly
from the ratio 
\bea
  \log 
  \frac {\cg_\Phi(\lambda,L)}{ \cg_\One(\lambda,L) }
&=& 
  \log\! 
  \frac {g_\Phi}{ g_\One }
-
  ( \lambda L^{12/5} )
  (2\pi)^{-7/5}
   I_1
   { \C\varphi\varphi\One  }
   \left( \fract 1{ \B\Phi\varphi\One}
   -  \fract 1{ \B\One\varphi\One} 
   \right)
  \;+\; \ldots
\;,
\eea 
so that \CPT\ predicts
\be
  c_{01} - d_1
=  (2\pi)^{-2/5}\,
   5^{-3/4}\,
   \frac{\Gamma(\fract 15)\, \Gamma(\fract 25)}
         {(1 + \sqrt 5)\,\Gamma(\fract 45)^2_{\vphantom\phi}}
\mathematica{
N[
    Pi^(-2/5) * 2^(-19/10) * 5^(-1/2) * Sin[2Pi/5]^(-1) *
   (1 + Sqrt[5])^(-1/2) * Gamma[1/5] * Gamma[2/5] * Gamma[4/5]^(-2)
]
N[
    Pi^(-2/5) * 2^(-2/5) * 5^(-3/4) * 
   (1 + Sqrt[5])^(-1) * Gamma[1/5] * Gamma[2/5] * Gamma[4/5]^(-2)
]
}
=  0.3328824..
\;.
\label{eq:c01d1}
\ee
In section \ref{sec:compare} we shall compare this with the TBA
results from section \ref{sec:tba0}.

\subsection{
The boundary perturbation
}
\label{sec:boundaryp}

The boundary condition $\Phi(h)$ 
corresponds to the perturbation of the $\Phi$
boundary by the addition to the action of the term
\be
  \delta S
= h \int \!|\dz |\,\phi(z)
\;.
\ee
To calculate $c_{10}$ and $c_{20}$ we can assume that the bulk theory
is unperturbed which means that we can replace the ground state
$\vec\Omega$ by $\vec\varphi$.
The perturbed boundary state is 
\[
  \cev{\Phi(h)}
= \cev{\Phi}
  \Big( 
  1 - \delta S + \fract 12 (\delta S)^2 + \ldots
  \Big)
\;,
\]
and so the coefficients in the expansion
\be
  \veev{\Phi(h)}{ \varphi }
=
  g_\Phi
\, \Big(
  1
\,+\, \tilde c_{10}\,  h L^{6/5}
\,+\, \tilde c_{20}\, (h L^{6/5})^2
\,+\, \ldots
   \Big)
\label{eq:ri}
\ee
can be written in terms of the simple disc amplitudes
\bea
    \tilde c_{10}
&=& 
 -  \frac{(2\pi)^{-6/5} }{g_\Phi}
    \int_{0}^{2\pi} 
    \Vev{\; \phi(e^{i\theta})\; \varphi(0) \;}^{\rm disc}_{\Phi} 
    \;\Dth
\;,
\nn\\
    \tilde c_{20}
&=& 
    \fract 12
    \frac{(2\pi)^{-12/5} }{g_\Phi}
    \int\limits_0^{2\pi}\!
    \int\limits_0^{2\pi}
    { \Vev{\; \phi(e^{i\theta})\;
                   \phi(e^{i\theta'})\; 
                   \varphi(0) \;}^{\rm disc}_{\Phi} }
    \;\Dth \, \Dth'
\;.
\label{eq:ctildes}
\eea
The correlation functions appearing in (\ref{eq:ctildes}) are
given in section \ref{sec:ylcfs}.
The two-point function
$\vev{\varphi(0)\phi(e^{i\theta})}^{\rm disc}_\Phi$
is rotationally invariant and so the integration in $\tilde c_{10}$ is
trivial, giving 
\be
  \tilde c_{10}
= - (2\pi)^{-1/5} 
  \frac{ \B \Phi\varphi\phi \C\phi\phi\One}{\B \Phi\varphi\One}
= (2\pi)^{-1/5} 5^{1/4}
  \left|
  \frac{ \Gamma(\fract 25) \Gamma( \fract 65 ) }
       { 2 \cos(\pi/5) \Gamma(\fract 45)^2     }
  \right|^{1/2}
= 0.99777..
\mathematica{0.997772821597 
 c1pcft = (2 Pi)^(-1/5) 5^(1/4) * 
          Sqrt[ Gamma[2/5] Gamma[6/5] / 
          ( 2 Cos[Pi/5] Gamma[4/5]^2 ) ]
}
\ee
It is a little more complicated to find $\tilde c_{20}$, which is
given as
\be
  \tilde c_{20}
= (2\pi)^{-7/5}
  \left\{
  \C\phi\phi\One\,
  I_3
\;+\;
  \C\phi\phi\phi\,
  \C\phi\phi\One\,
  \frac{ \B\Phi\varphi\phi}{\B\Phi\varphi\One}\,
  I_4
  \right\}
\;,
\label{eq:zp4}
\ee
where the integrals $I_3$ and $I_4$ are given by
\bea
  I_3
&=& \int_0^{\pi}
    (2\sin\fract\theta 2)^{2/5}
    (\cos\fract\theta 2)^{-4/5}
    F(\fract 4{10},\fract 9{10};\fract {11}{10};-\tan^2\!\fract\theta 2) 
    \; \Dth
\;,
\nn\\  
  I_4
&=& \int_0^{\pi}
    (2\sin\fract\theta 2)^{1/5}
    (\cos\fract\theta 2)^{-3/5}
    F(\fract 3{10},\fract 8{10};\fract 9{10};-\tan^2\!\fract\theta 2) 
    \; \Dth
\;, 
\eea
which do not seem to have simple closed expressions.
However, by making use of the substitution
$ v = \cos^2\!(\theta/2)$ 
and the explicit forms for the structure constants
appearing in (\ref{eq:zp4}), one can show that the specific
combination of $I_3$ and $I_4$ in (\ref{eq:zp4}) is given by 
\be
  \tilde c_{20}
= \frac{1}{\sqrt 5}\, ( \tilde c_{10} )^2
\;.
\label{eq:oh2}
\ee 
Taking the logarithm of (\ref{eq:ri}), we find 
\be
  \log \cg_{\Phi(h)}(0,L)
= \log g_\Phi
\;+\;
   \tilde c_{10} h L^{6/5}
\;+\;
   (\tilde c_{20} - \fract 12 (\tilde c_{10})^2)\,
   ( h L^{6/5})^2
\;+\;
\ldots
\;,
\ee
so that the coefficients $c_{10}$ and $c_{20}$ 
in (\ref{eq:cddefs}) are given by
\bea
  c_{10} 
&=& 
  (2\pi)^{-1/5} 5^{1/4}
  \left|
  \frac{ \Gamma(\fract 25) \Gamma( \fract 65 ) }
       { 2 \cos(\pi/5) \Gamma(\fract 45)^2     }
  \right|^{1/2}
= 0.99777..
\mathematica{0.997772821597 
 c1pcft = (2 Pi)^(-1/5) 5^(1/4) * 
          Sqrt[ Gamma[2/5] Gamma[6/5] / 
          ( 2 Cos[Pi/5] Gamma[4/5]^2 ) ]
}
\;,\;\;\;\;
\nn\\
  c_{20} 
&=&
  \left( \fract1{\sqrt 5} {-}\fract 12 \right)
  (c_{10})^2
\;.
\label{eq:c10c20}
\eea
The coefficients $c_{10}$ and $c_{20}$ have also been calculated using
the free-field construction in \cite{BLZ2}, with the same results that
we find.

\subsection{Evaluating ${\log(g_{\alpha})}$ using the TCSA method}
\label{sec:btcsa}

The truncated conformal space approximation (TCSA) 
introduced in \cite{YZam1}
gives numerical
approximations to the lowest eigenvalues of the Hamiltonian of a
perturbed conformal field theory by restricting the Hamiltonian to a
finite dimensional subspace, working out the matrix elements exactly and
diagonalising the resulting matrix numerically. 

Originally developed for systems on a circle, we outlined in 
\cite{Us1} how this method can also be applied to a system on a 
strip with conformal boundary conditions $\alpha$, $\beta$
on the two ends and perturbed simultaneously by
a bulk field $\Phi$ of weight $x_\Phi$ 
and by boundary fields $\phi_l$ and $\phi_r$ of
weights $x_l$ and $x_r$.
The resulting Hamiltonian is
\[
\begin{array}{rl}
&
  \hat H(R,\lambda,h_l,h_r)
\\
= 
& \ds
\frac{\pi}{R} 
    \Big(
  L_0 - \fract{c}{24}
\;\;+\;\;
    \lambda 
    \left| \fract{R}{\pi} \right|^{2 - x_\Phi}
    \!\!\int\limits_{\theta = 0}^\pi \!\!
    \Phi(\exp(i\theta))
    \,\D\theta
\;\;+\;\;
     h_l
    \left| \fract{R}{\pi} \right|^{1 - x_l}
    \phi_l(-1)
\;\;+\;\;
    h_r
    \left| \fract{R}{\pi} \right|^{1 - x_r}
    \phi_r(1)
  \Big)
\;.
\end{array}
\]
The matrix elements of $\hat H$ can be easily calculated numerically
and the Hamiltonian diagonalised on spaces of the
order of 100 states.

The partition function on a cylinder of circumference $L$ can  be
estimated numerically as the trace over the finite-dimensional
truncated space.
\[
  Z_{\alpha(h_l)\beta(h_r)}(L,R,\lambda)
= \Tr( \exp( - L \hat H(R,\lambda,h_l,h_r)))
\;.
\]
More details will be given in the next section where the method is
applied to perturbations of the  $\One$ and $\Phi(0)$
boundary conditions, both by bulk perturbations, and (for the
$\Phi(0)$ boundary condition) at the boundary.
In the presence of the bulk perturbation, the mass scale is set by the
mass $M$ of the single massive particle of the scaling Yang--Lee
model,
which is related to $\lambda$ by
\cite{Zb,Zg}
\[
  M 
= \kappa \lambda^{5/12} 
\;,
\;\;\;\;
  \kappa
= 2^{19/12}\sqrt{\pi}\,
  \ds\frac
  {  ~\left( \Ga(3/5)\Ga(4/5) \right)^{5/12} }
  {   5^{5/16}\Ga(2/3)\Ga(5/6)  }
= 2.642944\dots
\;,
\]
We present the results for the $(\One,\One)$ and $(\One,\Phi(h))$
boundary conditions in the next two sections.

\subsubsection{The strip with ${(\One,\One)}$ boundary conditions}
\label{sec:oo}

The $(\One,\One)$ system has no relevant boundary perturbations and
so the strip Hamiltonian is simply
\[
  \hat H(R,\lambda)
= 
\frac{\pi}{R} 
    \Big(
  L_0 - \fract{c}{24}
\;\;+\;\;
    \lambda 
    \left| \fract{R}{\pi} \right|^{2 - x_\Phi}
    \!\!\int\limits_{\theta = 0}^\pi \!\!
    \Phi(\exp(i\theta))
    \,\D\theta
  \Big)
\;.
\]
The spectrum comprises scattering states of a single massive
particle of mass $M$ and no boundary bound states.
In figure \ref{fig:Graph1} we plot the 15 lowest eigenvalues of 
$\hat H/M$ against $r = MR$ (as estimated from TCSA with 29 states).

We estimate the partition function 
by taking the sum over all the TCSA eigenvalues $E_i^{\rm TCSA}$
at the given truncation level
\[
   Z(r,l) 
= \sum_{i} \exp( - L E^{\rm TCSA}_i )
\;,
\]
where $l = M L$.
For any particular $l$, we expect that the behaviour of
$Z(r,l)$ for small $r$ is dominated by the conformal point, while at
large $r$ truncation errors will dominate. If we are lucky, there will
be an intermediate range of $r$ (the scaling region) in which $Z(r,l)$
exhibits the scaling form
\be
     \log Z(r,l)
\sim -a(l)\, r + b(l) + \ldots
\;.
\label{eq:scale}
\ee
{}From equations (\ref{lrasympt}), (\ref{eq:aml}) and
(\ref{eq:cgdef}), we expect that
\[
  a(l) = \frac{ E_0^{\rm cyl}}M
\;,\;\;\;\;
  b(l) = 2 \log \cg_\One(l)
\;.
\]
To test for the behaviour (\ref{eq:scale}) we plot
$ \log( \D \log Z(r,l) / \D r ) $ to see if, for given $l$,  there is
a region in $r$ for which it is approximately constant. 
In figure \ref{fig:Graph2} we plot
$ \log( \D \log Z(r,l) / \D r ) $ against $\log(r/l)$ for
various fixed values of $l$ calculated using the TCSA with 19 states
and with 106 states.
For small values of $l$ we see the typical behaviour expected --  
as $r$ increases, $ \log( \D \log Z(r,l) / \D r ) $ 
flattens out to a scaling region in which it is approximately constant
before truncation effects take over.
We can be fairly sure that this is indeed the scaling region by
comparing the behaviour for different truncation levels, and, as
expected, increasing the truncation level extends the scaling region
further into the IR.
On close examination we find for small $l\lesssim 1$ that the scaling
region  is centred on 
$\log(r/l) \sim 1.21$ for 19 states
and on 
$\log(r/l) \sim 1.46$ for 106 states.
Hence we shall take the TCSA estimates of $a(l)$ and $b(l)$ to be those
at 
$\log(r/l) = 1.21$ 
and 
$\log(r/l) = 1.46$ 
for 19 and 106 states respectively.
For larger values of $l> 1$ the effects of truncation start to
interfere for $\log r \sim 2.75$ and it is not clear whether we have
actually reached the scaling region or not. 

\[
\begin{array}{cc}
\refstepcounter{figure}
\label{fig:Graph1}
\refstepcounter{figure}
\label{fig:Graph2}
\epsfxsize=.45\linewidth
\epsfbox[0 50 288 238]{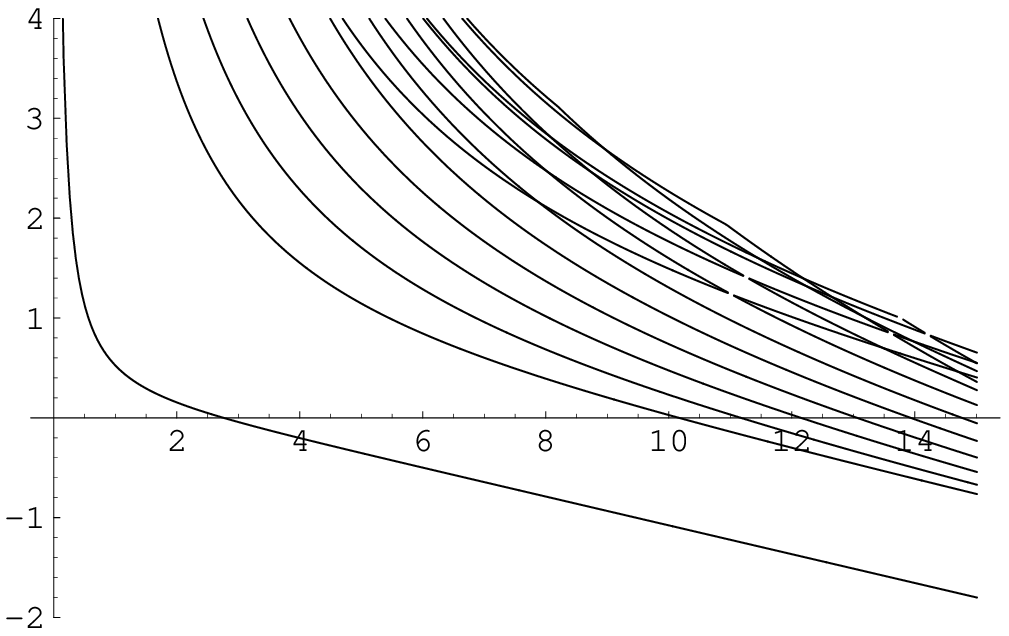}
&
\epsfxsize= .45\linewidth
\epsfbox[0 50 288 238]{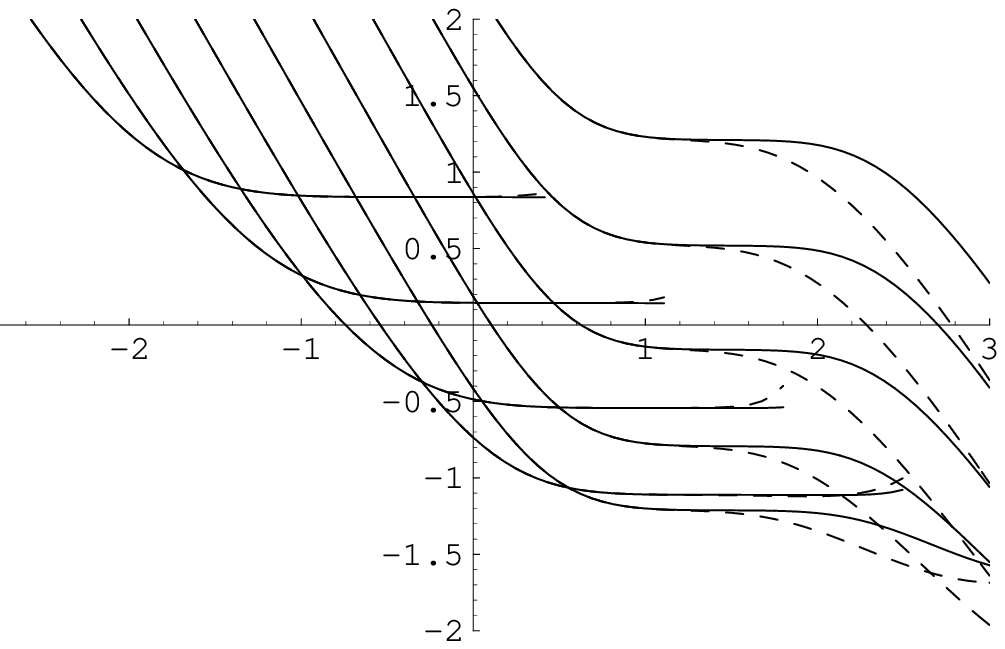}
\\
\parbox[t]{.48\linewidth}{\small\raggedright%
Figure \ref{fig:Graph1}:
The first 15 eigenvalues of
$\hat H/M$ for the $(\One,\One)$ system 
plotted against $r$ from 
TCSA with 29 states. 
}
&
\parbox[t]{.48\linewidth}{\small\raggedright%
Figure \ref{fig:Graph2}:
Plots of $ \log( \D \log Z(r,l) / \D r ) $ vs.\ $\log(r/l)$ 
for $l {=} 2^n$, with  (from left to right)
$n = 4,3,2,1,0,-1,-2,-3,-4$.
Results are for 106 (solid) and for 19 (dashed) states.
}
\end{array}
\]
In figure \ref{fig:Graph3} we plot $a(l)$ as estimated in this way
from the TCSA with 106 states
and include a plot of
$E_0^{\rm circ}(l)/M$ 
for comparison. It is clear from the graph
that it is possible to estimate $a(l)$ effectively this way.

In figure \ref{fig:Graph4} we plot $b(l)$ as estimated in the same
way. The onset of the linear behaviour (\ref{logg}) expected for
large $l$ is clear in this plot. 
In figure \ref{fig:Graph5} we plot  $\D b(l)/\D l$ and give for
comparison the exact TBA prediction from eqn.\ (\ref{eq:ftba}). It is 
clear that while the TCSA cannot predict the correct 
value, it agrees with the TBA 
to our accuracy, and so in figure
\ref{fig:Graph6} we plot the combination 
\[
  L f^{\rm TBA}_\One
+
  \log \cg_\One^{\rm TCSA}(l) 
\;,\;\;\;\;
  f^{\rm TBA}_\One = \fract 14 (\sqrt 3 - 1)
\;,
\]
showing its interpolation between the UV and IR
values.  
Finally in table \ref{tab:Table1}
we give the TCSA estimate of the coefficients $\cg^{(n)}(0) / n!$
of the series expansion (\ref{eq:cmndn}) from truncations to 
finite dimensional spaces and the extrapolation 
of these coefficients to infinite level: 
%
%
%
%
\refstepcounter{table}
\label{tab:Table1}
\[
  \begin{array}{c|lllll|l}
      & \multicolumn{5}{c|}{\rm TCSA} 
      & 
   \\
 
      & \hbox{63 states}  
      & \hbox{75 states}  
      & \hbox{90 states}  
      & \hbox{106 states}  

      & \hbox{$\infty$ states}

      & \multicolumn{1}{c }{\rm exact} 
    \\ \hline

 d_0  

      & -0.321615 
      & -0.321558 
      & -0.321518 
      & -0.321580 

      & -0.3215     

      & -0.32148269..  \\

 d_1  
      & -0.25230 
      & -0.25252 
      & -0.25269 
      & -0.25260 

      & -0.253   
      
      & -0.25311758.. \\

 d_2  
      & \m 0.076715 
      & \m 0.077208 
      & \m 0.077523 
      & \m 0.077570 

      & \m 0.0775  

      &  \multicolumn{1}{c}{\hbox{---}} \\

 d_3  
      & -0.03510 
      & -0.03569 
      & -0.03599 
      & -0.03606 

      & -0.0360  

      &   \multicolumn{1}{c}{\hbox{---}} \\

 d_4

      & \m 0.01787 
      & \m 0.01841 
      & \m 0.01857 
      & \m 0.01893 

      & \m 0.019

      &   \multicolumn{1}{c}{\hbox{---}} \\

\multicolumn{7}{c}{\hbox{Table \ref{tab:Table1}}}

  \end{array}
\]
\vskip -5mm
\[
\begin{array}{cc}
\refstepcounter{figure}
\label{fig:Graph3}
\refstepcounter{figure}
\label{fig:Graph4}
\epsfxsize=.45\linewidth
\epsfbox[0 50 288 238]{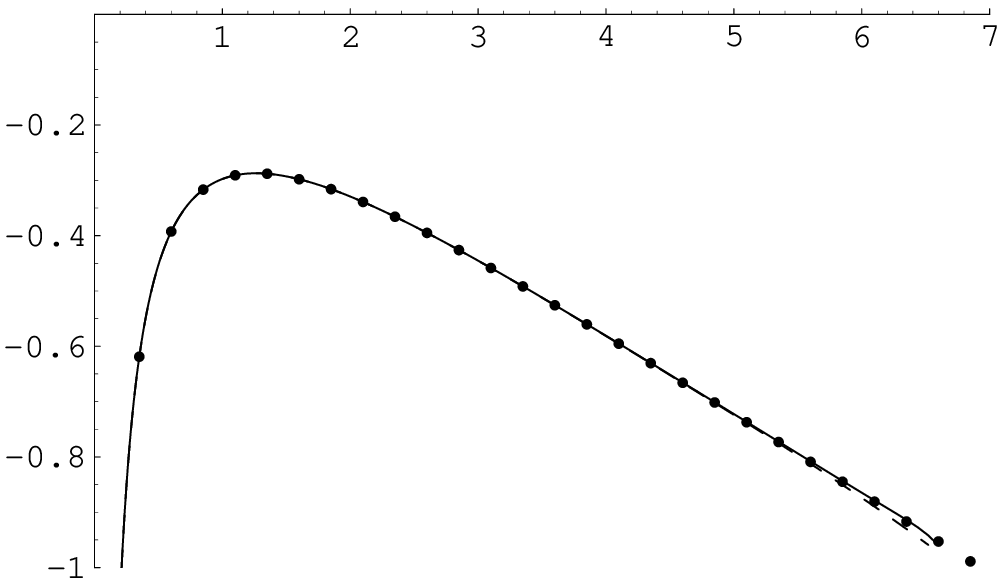}
&
\epsfxsize= .45\linewidth
\epsfbox[0 50 288 238]{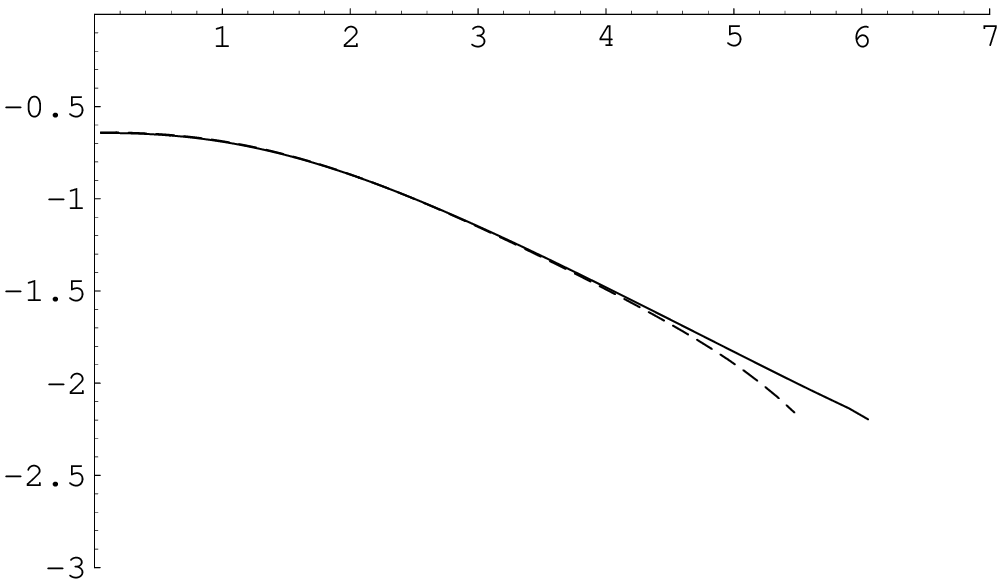}
\\
\parbox[t]{.45\linewidth}{\small\raggedright%
Figure \ref{fig:Graph3}:
$a(l)$ as estimated from TCSA with 19 states (dashed line) and 
106 states (solid line) together with 
$E_0^{\rm circ}(l)/M$ (points) plotted against $l$.
}
&
\parbox[t]{.45\linewidth}{\small\raggedright%
Figure \ref{fig:Graph4}:
$b(l)$ as estimated from TCSA with 19 states (dashed line) and 
106 states (solid line)
plotted against $l$.
}
\\
\refstepcounter{figure}
\label{fig:Graph5}
\refstepcounter{figure}
\label{fig:Graph6}
\epsfxsize=.45\linewidth
\epsfbox[0 50 288 238]{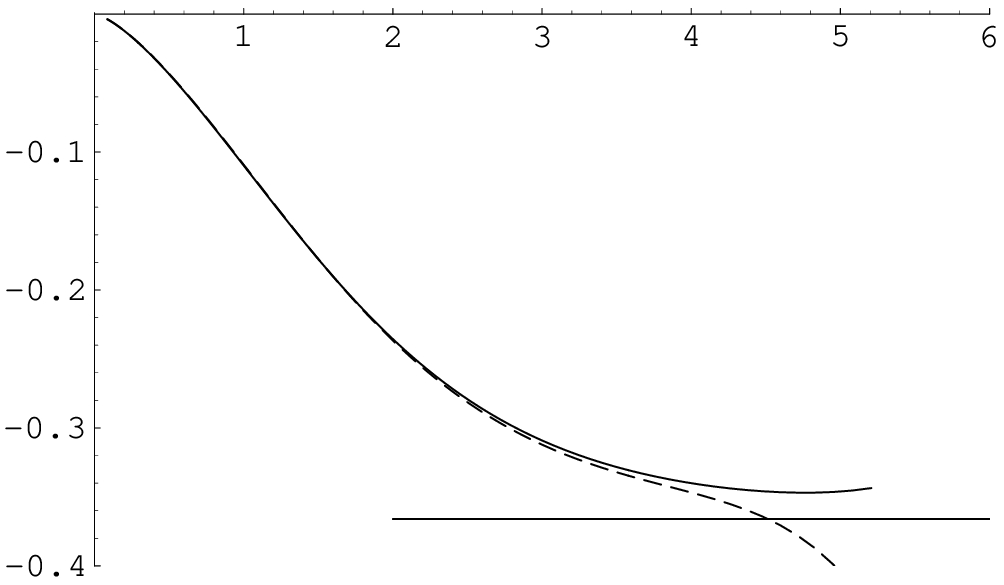}
&
\epsfxsize= .45\linewidth
\epsfbox[0 50 288 238]{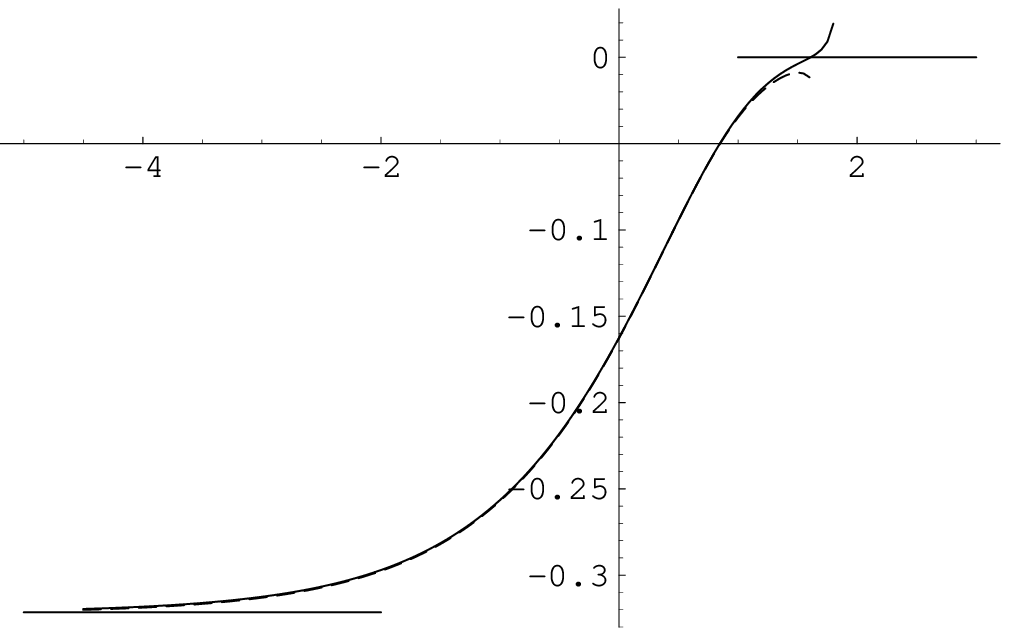}
\\
\parbox[t]{.45\linewidth}{\small\raggedright%
Figure \ref{fig:Graph5}:
$\D b/\D l$ as estimated from TCSA with 
19 states (dashed line) and
106 states (solid line) plotted against $l$.
Also shown is the TBA prediction 
$- 2 f^{TBA}_\One$. 
}
&
\parbox[t]{.45\linewidth}{\small\raggedright%
Figure \ref{fig:Graph6}:
$\log g_\One(l)$ vs.\ $\log l$
from $f^{\rm TBA}_\One$ and the 
\cgf\ from the TCSA with 
19 (dashed line) and
106 (solid line) states. Also shown are the exact UV and IR values. 
}
\end{array}
\]

\subsubsection{The strip with $(\One,\Phi(h))$ boundary conditions:
massive case}
\label{sec:op}

We now consider the strip with boundary conditions $(\One,\Phi(h))$,
corresponding to a Hamiltonian
\be
  \hat H(R,\lambda,h)
= 
\frac{\pi}{R} 
    \Big(
  L_0 - \fract{c}{24}
\;\;+\;\;
    \lambda 
    \left| \fract{R}{\pi} \right|^{12/5}
    \!\!\int\limits_{\theta = 0}^\pi \!\!
    \varphi(\exp(i\theta))
    \,\D\theta
\;\;+\;\;
    h
    \left| \fract{R}{\pi} \right|^{6/5}
    \phi(1)
  \Big)
\;.
\label{eq:stripham}
\ee
The general partition function is now a function of three dimensionless
combinations of $R$, $L$, $\lambda$ and $h$.
In  the massive case it is convenient to take these as the
scaled lengths and the scaled boundary field, defined by 
\be
   r = M \, R
\;,\;\;\;\;
   l = M \, L
\;,\;\;\;\;
   \hat h = h \, M^{-6/5}
\;,
\label{eq:param1}
\ee
with the expectation that 
\bea
  \log Z_{(\One,\Phi(h))}(R,L,\lambda) 
&\sim& 
  - r\,a(l,\hat h) 
  \;+\;
   b(l,\hat h)
\;,
\label{eq:B1}
\\[2mm]
  b(l,\hat h)
= 
  \log(\, \cg_\One(L,\lambda) \, \cg_{\Phi(h)}(L,\lambda) \,)
&\sim& 
- {\ds\frac lM}
  \,( f_{\Phi(\hat h)} + f_\One )
  \;+\;
  \log(\, g_\One(l) \,  g_{\Phi(\hat h)}(l) \,)
\;.~~~
\label{eq:B2}
\eea
To simplify matters, in this section we will only discuss the model
with a purely bulk perturbation, so that 
$h=0$. This corresponds to the value $b=-1/2$ of the IR boundary
parameter defined in eqn.\ (\ref{eq:hnum}).
In figure \ref{fig:Graph7} we show the spectrum of the model with
these boundary conditions and
truncation to 81 states (the first excited state here is actually a
boundary bound state). 
Examining $\D \log(Z(r,l))/\D r$, we find that, as before,
for small values of $l$ there is a scaling region in $r$   
which broadens with increasing $l$ before truncation effects intervene.
For 140 states, the small $l$ scaling region is centred on
$\log(r/l) \sim 1.38$ while for large $l$ truncation effects occur for
$\log r \sim 1.5$. 
For small $l$ we can clearly see that scaling has set in and can be
confident that we can estimate $a(l)$ and $b(l)$ using the values at 
the midpoint of the scaling region. For large $l$ we can assume that
the `best guesses' of $a(l)$ and $b(l)$ are given by their 
values at either
(a) the small $l$ effective value of $r$, or 
(b) the largest
value of $r$ before truncation effects are obvious. In other words we
consider the two estimates of $r_{\rm eff}(l)$ defined by
\[
(a)\;
  \log r_{\rm eff}(l) 
= 1.38 + \log l
\;,\;\;\;\;
\hbox{ and }
(b)\;
  \log r_{\rm eff}(l) 
= \min(\;1.38 + \log l\;,\; 1.5\;)
\;\;.
\]

In figures \ref{fig:Graph9} and \ref{fig:Graph11}
we plot $a(l)$ and $\D b/\D l$ 
using these two methods. We see that this second method of taking the
`best guess' at the large $l$ values gives quite pleasing results 
for these two quantities
while having the problem that the errors are hard to estimate.
In figure \ref{fig:Graph12} we 
plot $\log(g_\One(l)\, g_{\Phi(0)}(l))$
using method (b) and compare the
results using the TBA value of $f_\alpha$ and a `fitted value' as was
used in \cite{Us1}. As we see, 
the large $l$ behaviour of the \gf\ appears to be improved by using a
`fitted' value of $f_\alpha$, but again
the errors induced by doing this are unknown.

\[
\begin{array}{cc}
\refstepcounter{figure}
\label{fig:Graph7}
\refstepcounter{figure}
\label{fig:Graph9}
\epsfxsize=.45\linewidth
\epsfbox[0 50 288 238]{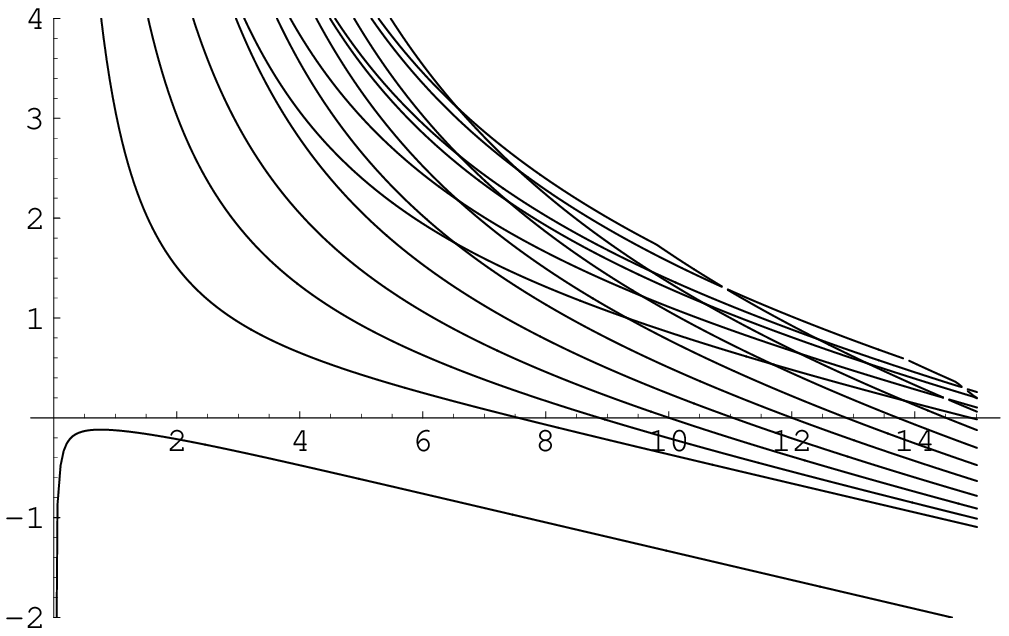}
&
\epsfxsize= .45\linewidth
\epsfbox[0 50 288 238]{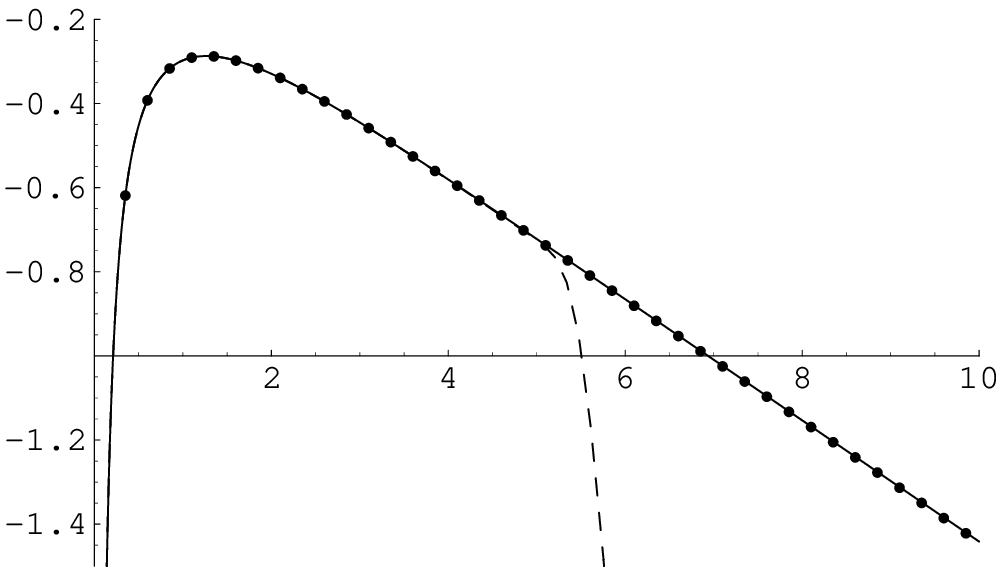}
\\
\parbox[t]{.45\linewidth}{\small\raggedright%
Figure \ref{fig:Graph7}:
The first 15 eigenvalues of
$\hat H/m$ for the $(\One,\Phi(0))$ system 
plotted against $r$ from 
TCSA with 81 states. 
}
&
\parbox[t]{.45\linewidth}{\small\raggedright%
Figure \ref{fig:Graph9}:
$a(l)$ vs.\ $l$, as estimated from TCSA with 140 states using methods
(a) (dashed line) and 
(b) (solid line), together with 
$E_0^{\rm circ}(l)/M$ (points).
}
\end{array}
\]
\[
\begin{array}{cc}
\refstepcounter{figure}
\label{fig:Graph11}
\refstepcounter{figure}
\label{fig:Graph12}
\epsfxsize=.45\linewidth
\epsfbox[0 50 288 238]{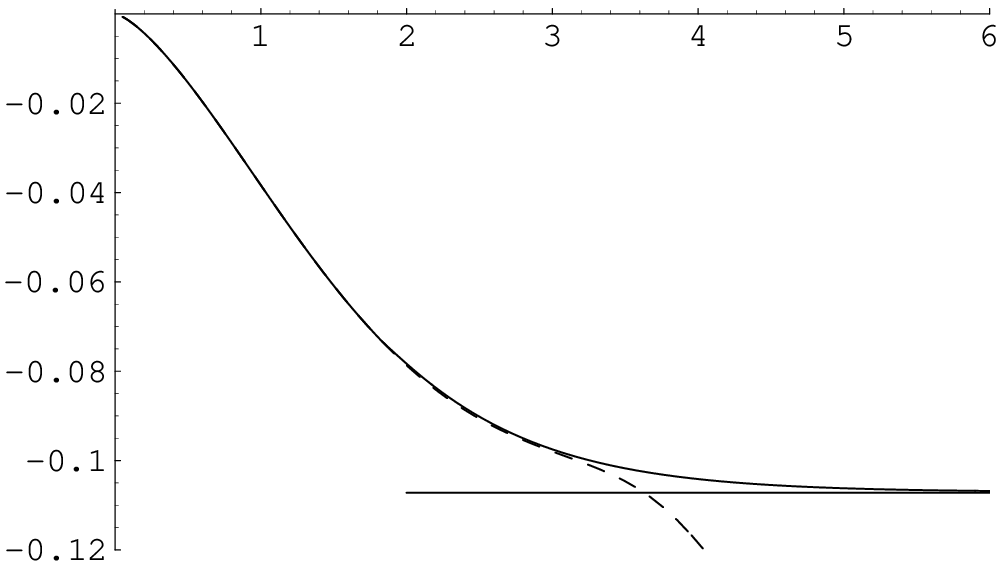}
&
\epsfxsize= .45\linewidth
\epsfbox[0 50 288 238]{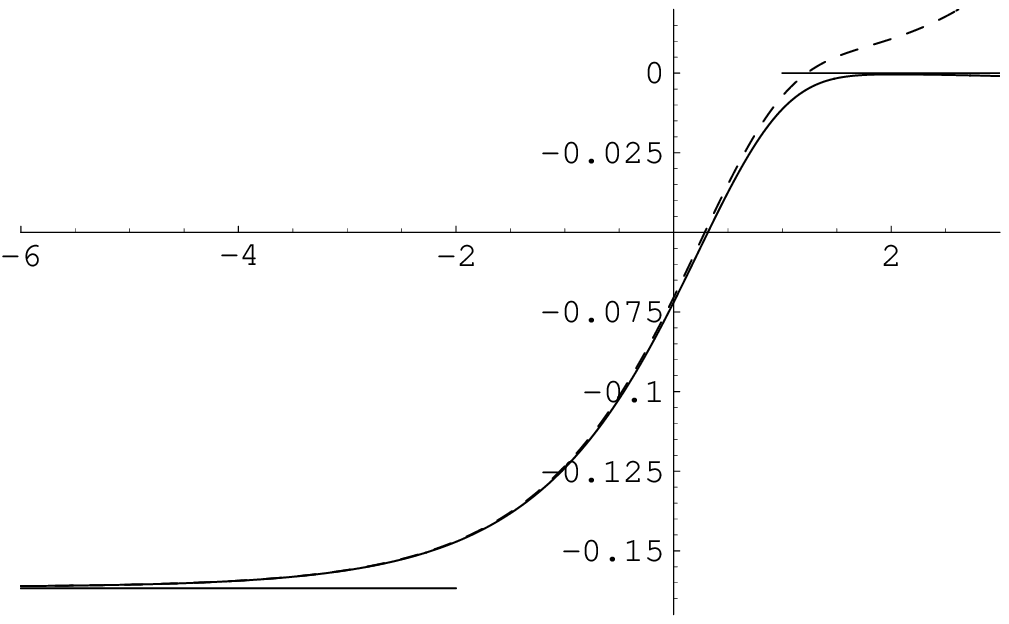}
\\
\parbox[t]{.45\linewidth}{\small\raggedright%
Figure \ref{fig:Graph11}:
$\D b/\D l$ as estimated from TCSA with 
with 140 states using methods
(a) (dashed line) and 
(b) (solid line), together with 
the TBA prediction.
}
&
\parbox[t]{.45\linewidth}{\small\raggedright%
Figure \ref{fig:Graph12}:
$\log( g_\One(l)\, g_{\Phi(0)}(l))$ 
vs.\ $\log l$
from the TCSA with 140 states using method (b)
and 
$f^{\rm TBA}_\alpha$ (dashed line) and a fitted value of
$f_\alpha$ (solid line). 
Also shown are the exact UV and IR values. 
}
\end{array}
\]

In table \ref{tab:Table2} we give the coefficients in the power series
expansion (\ref{eq:cmndn}) as found from the TCSA data by estimating
the $n$-th derivatives of  $\log( \cg_\One \cg_{\Phi(0)})$ wrt
$\lambda$ at $\lambda=0$. We also give the result of extrapolating
these coefficients to infinite level. 

Combined with the results for $d_m$ in table \ref{tab:Table1} we can
then estimate the coefficients $(c_{0m}-d_m)$ which appear in the
expansion of the ratio $\log(\cg_{\Phi(h)} / \cg_\One)$.
These are compared with the results from the TBA 
in section \ref{sec:compare}, table {\ref{tab:Table2b}}.

\refstepcounter{table}
\label{tab:Table2}
\[
  \begin{array}{c|lllll|l}
      & \multicolumn{5}{c|}{\rm TCSA} 
      & \multicolumn{1}{c}{\rm exact} 
 \\
 

      & \hbox{81 states}  
      & \hbox{98 states}  
      & \hbox{117 states}  
      & \hbox{140 states}  

      & \hbox{$\infty$ states}

      & 
   \\ \hline

\blank{
10   {-0.15837964, -0.17943884, 0.069252021, -0.038275552, 0.023022875}
11   {-0.15921615, -0.17813719, 0.069134682, -0.038685104, 0.02321856}
12   {-0.15985831, -0.17714128, 0.068470197, -0.038303042, 0.023350031}
13   {-0.1603091,  -0.17629895, 0.068078045, -0.037949011, 0.022954339}
14   {-0.16064396, -0.17578295, 0.067479549, -0.037475599, 0.022937879}
15   {-0.16089084, -0.17534168, 0.067446257, -0.037044537, 0.022481385}
16   {-0.16108352, -0.17499932, 0.06702825,  -0.036968186, 0.022669219}
17   {-0.16122492, -0.1747552,  0.06712173,  -0.03658558,  0.022298421}
18   {-0.16133749, -0.17456754, 0.066757201, -0.036288861, 0.022254023}
}

 c_{00} + d_0 
     & -0.16089 
     & -0.16108 
     & -0.16123 
     & -0.1613 

     & -0.162

     & -0.161753565..
   \\

 c_{01} + d_1 

     & -0.1754 
     & -0.1750 
     & -0.1748 
     & -0.1746 

     & -0.174  

     &  -0.173352755..

   \\

 c_{02} + d_2 

     & \m 0.06745 
     & \m 0.06703 
     & \m 0.06712 
     & \m 0.06676 

     & \m 0.0628 

     &  \multicolumn{1}{c}{\hbox{---}}
   \\

 c_{03} + d_3 

     & -0.0370 
     & -0.0370 
     & -0.0366 
     & -0.0363 

     &  -0.033 

     &  \multicolumn{1}{c}{\hbox{---}}
   \\

 c_{04} + d_4

& \m 0.02248 
& \m 0.02267 
& \m 0.02230 
& \m 0.02225 

     &  \m 0.02 

     &  \multicolumn{1}{c}{\hbox{---}}
   \\

  \end{array}
\]
\centerline{%
Table \ref{tab:Table2}:
The coefficients $(c_{0n}{+}d_n)$ from TCSA and \CPT
}

\subsubsection{ $(\One,\Phi(h))$ boundary conditions: massless case}

We now consider the $(\One,\Phi(h))$ model with the bulk massless. For
this case, the
three combinations (\ref{eq:param1}) do not make much sense, and it
is better to consider the system as a function of the two variables
\be
  \tilde r 
= r/l = R/L
\;,\;\;\;\;
  \tilde h 
= \hat h \, l^{6/5} = h \, L^{6/5}
\;.
\label{eq:param2}
\ee
As was shown in figure 1(b) of our previous paper \cite{Us1}, 
the spectrum 
of $\hat H$ here is
real for $\tilde h$ positive, and for $\tilde h \gg 1$  approaches that
of the strip with $(\One,\One)$ boundary conditions. For sufficiently
negative $\tilde h$, the spectrum becomes complex.

The Hamiltonian on the circle is unperturbed if the bulk is massless, 
and
so the ground state energy on the circle is simply
given by (\ref{eq:cylham}):
\[
  R\, E^{\rm circ}_0(0)
= -\frac \pi{15}\,\frac RL
= - \frac{\pi}{15}\,\tilde r
\;.
\]
Similarly the massless limit of the linear term in $\cg$,
$(L f_b)$, is simply given by (\ref{eq:fb2}), 
so that (\ref{eq:B1}) and (\ref{eq:B2}) become
\bea
  \log Z_{(\One,\Phi(h))}(R,L,0)
&\sim&
  \frac{\pi}{15}\,\tilde r 
  \;+\; 
  b(\tilde h)
\;,
\label{eq:B3}
\\[2mm]
  b(\tilde h)
&\sim&
  - 2^{-1/6} \,
  \left|\,\frac {\tilde h}{h_c}\,\right|^{5/6}
  + \log( g_{\Phi(h)}(L,0) )
\;.
\label{eq:B4}
\eea

In figures \ref{fig:Graph14}a--\ref{fig:Graph14}c we show the TCSA
estimates of $Z(\tilde r,\tilde h)$ for various values of $\tilde h$ 
for truncations to 29, 67 and 140 states.
There is good convergence for $\tilde r \lesssim 8$.
\[
\begin{array}{ccc}
\refstepcounter{figure}
\label{fig:Graph14}
\epsfxsize=.32\linewidth
\epsfbox[0 50 288 238]{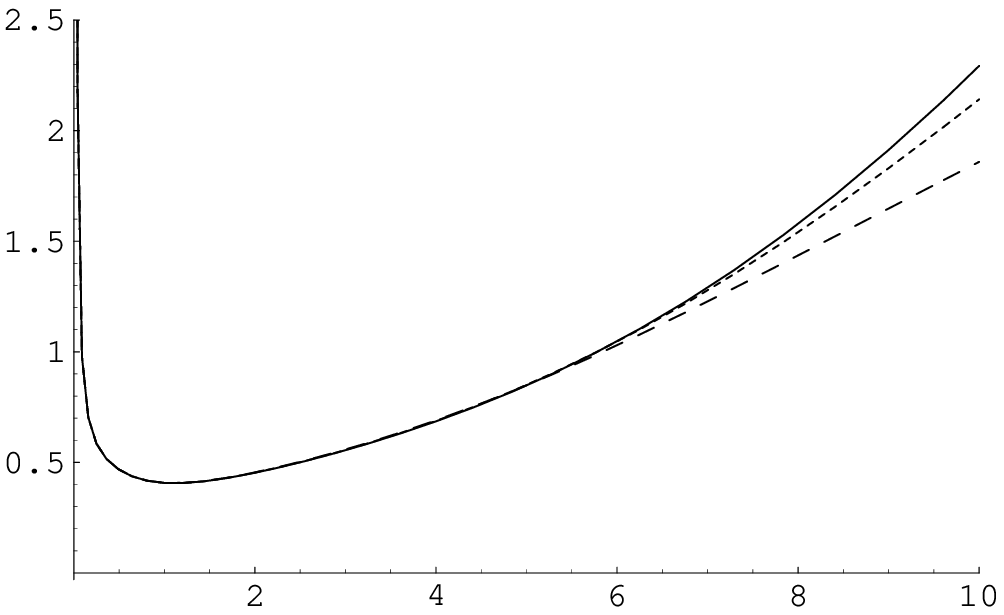} 
&\epsfxsize=.32\linewidth
\epsfbox[0 50 288 238]{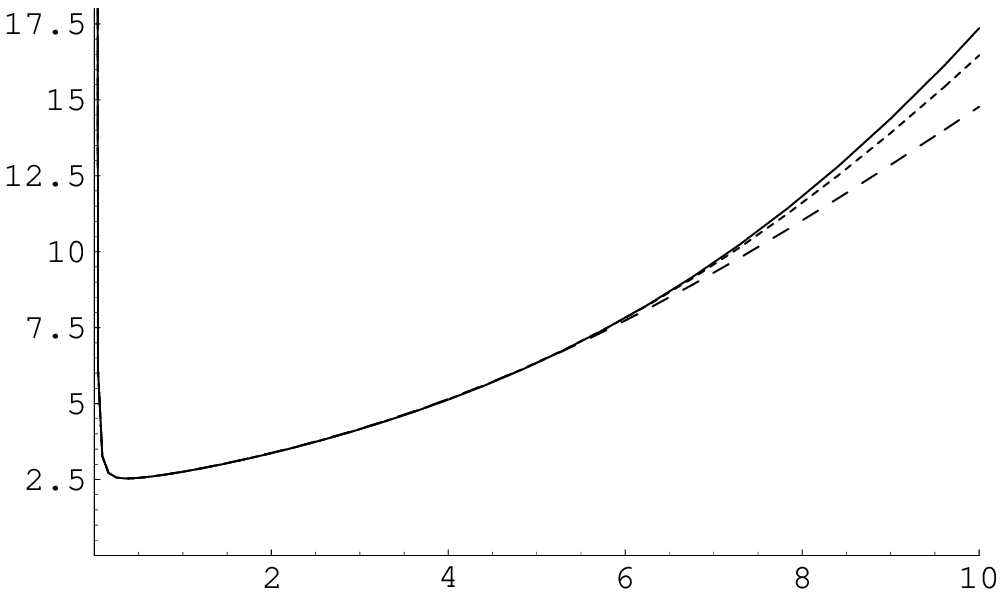} 
&\epsfxsize=.32\linewidth
\epsfbox[0 50 288 238]{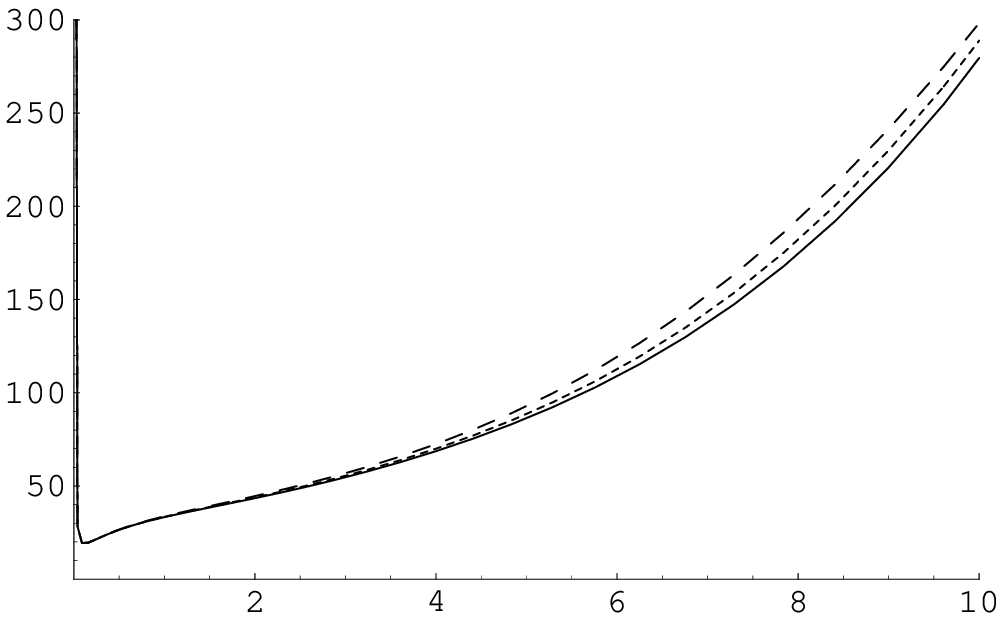} 
\\
{\rm (a)}\;\;\;\; \tilde h = -1\;.
&
{\rm (b)}\;\;\;\; \tilde h = 1\;.
&
{\rm (c)}\;\;\;\; \tilde h = 4\;.
\\
\multicolumn{3}{c}{
\parbox[t]{.75\linewidth}{\small\raggedright%
Figure \ref{fig:Graph14}:
Graphs of $\log Z$ vs.\ $\tilde r$ for fixed values of $\tilde h$ 
from the TCSA to 29 states (dashed line), 67 states (dotted line) and
140 states (solid line)
}}
\end{array}
\]
In figures \ref{fig:Graph15}a--\ref{fig:Graph15}c we plot
$\log( \D \log( Z(\tilde r,\tilde h))/\D \tilde r)$ vs.\ 
$\tilde r$ to show the scaling region in which 
$\log Z$ grows approximately linearly, and include the exact value 
$\log(\pi/15)$ from eqn.\ (\ref{eq:B3}) for comparison.
We see from these graphs that for small values of $\tilde h$ there is
good convergence for $\tilde r \lesssim 3-6$.
{
\[
\begin{array}{ccc}
\refstepcounter{figure}
\label{fig:Graph15}
\epsfxsize=.32\linewidth
\epsfbox[0 50 288 238]{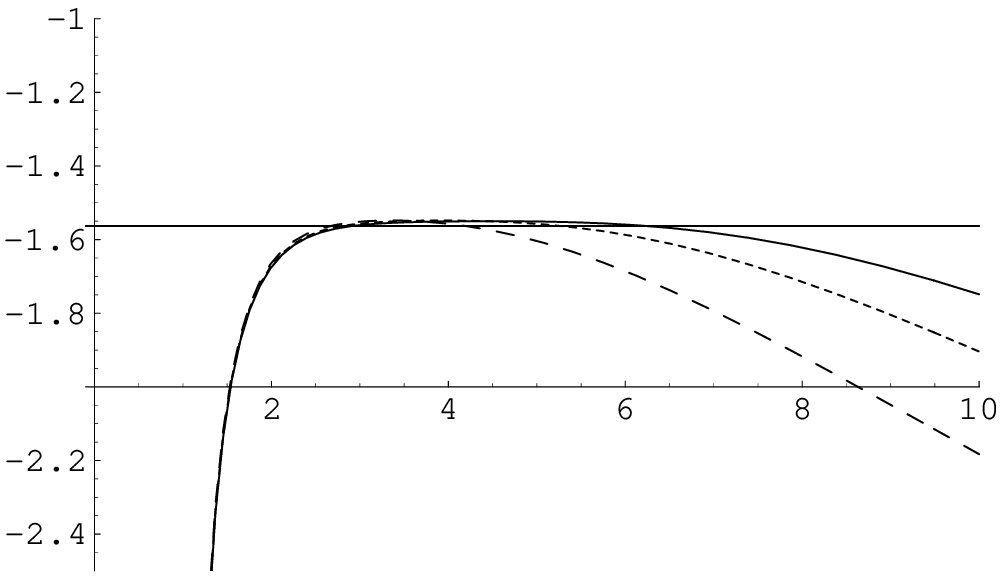} 
&
\epsfxsize=.32\linewidth
\epsfbox[0 50 288 238]{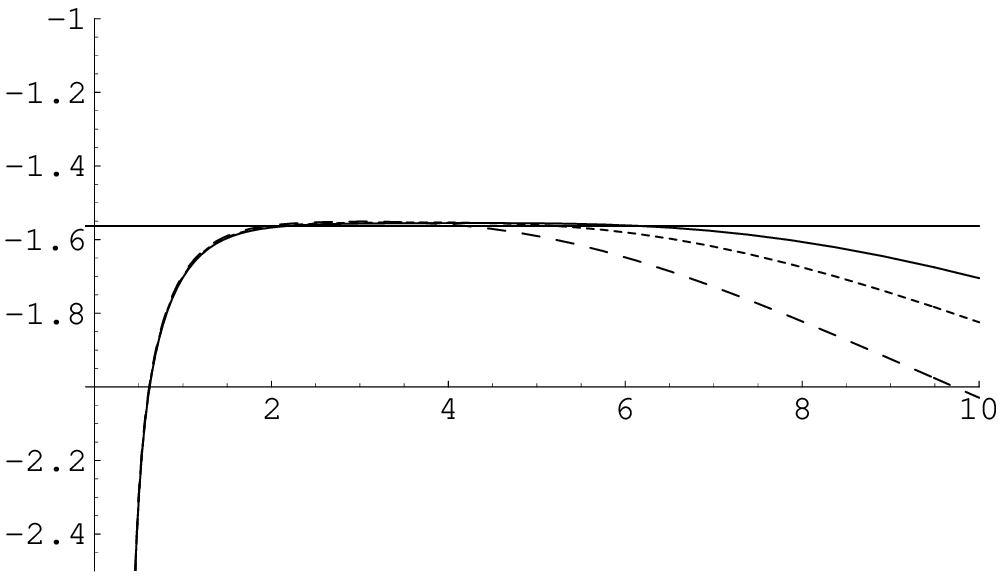} 
&
\epsfxsize=.32\linewidth
\epsfbox[0 50 288 238]{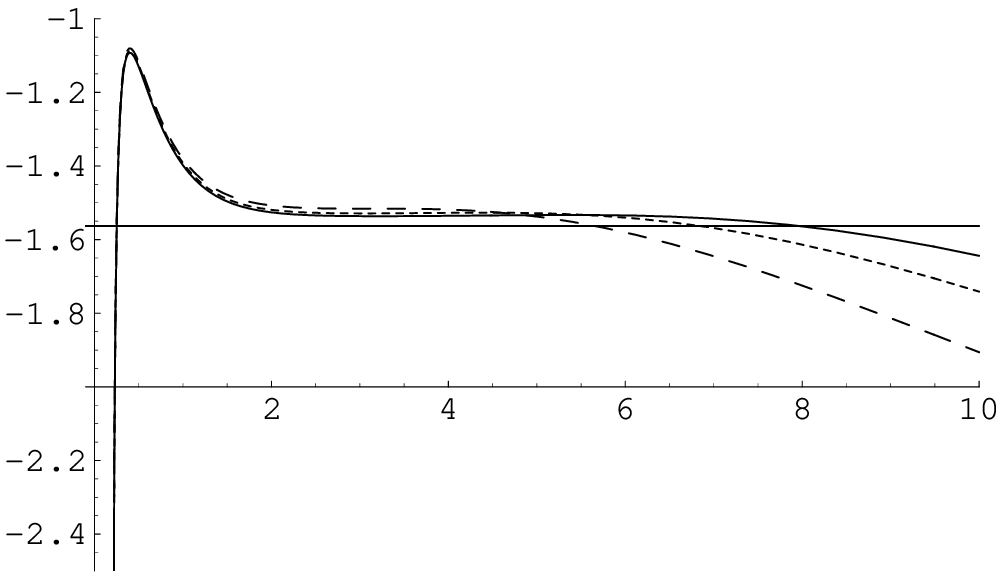} 
\\
{\rm (a)}\;\;\;\; \tilde h = -1\;.
&
{\rm (b)}\;\;\;\; \tilde h = 1\;.
&
{\rm (c)}\;\;\;\; \tilde h = 2\;.
\\
\multicolumn{3}{c}{
\parbox[t]{.75\linewidth}{\small\raggedright%
Figure \ref{fig:Graph15}:
Graphs of $\log(\D \log Z/\D\tilde r)$ vs.\ $\tilde r$ for fixed
values of $\tilde h$  
from the TCSA to 29 states (dashed line), 67 states (dotted line) and
140 states (solid line).
Also shown is the exact value $\log(\pi/15)$.
}}
\end{array}
\]
}
We can estimate $b(\tilde h)$ in two ways -- using the exact value of 
$ R E_0^{\rm circ} $ or using the TCSA value calculated using eqn.\
(\ref{eq:B2}). 
With $b(\tilde h)$ calculated in either manner we can finally use the
TBA value of $(f L)$ to calculate the \gf\ from eqn.\ (\ref{eq:B4}).
In figure \ref{fig:Graph16} we plot the TCSA estimate of 
$ \log(g_\Phi(h)) $
using the exact value of $E_0^{\rm circ}$ and the estimated value,
together with the massless TBA result (suitably normalised).
We see that there is excellent agreement.
{
\[
\begin{array}{c}
\refstepcounter{figure}
\label{fig:Graph16}
\epsfxsize=.48\linewidth
\epsfbox[0 50 288 238]{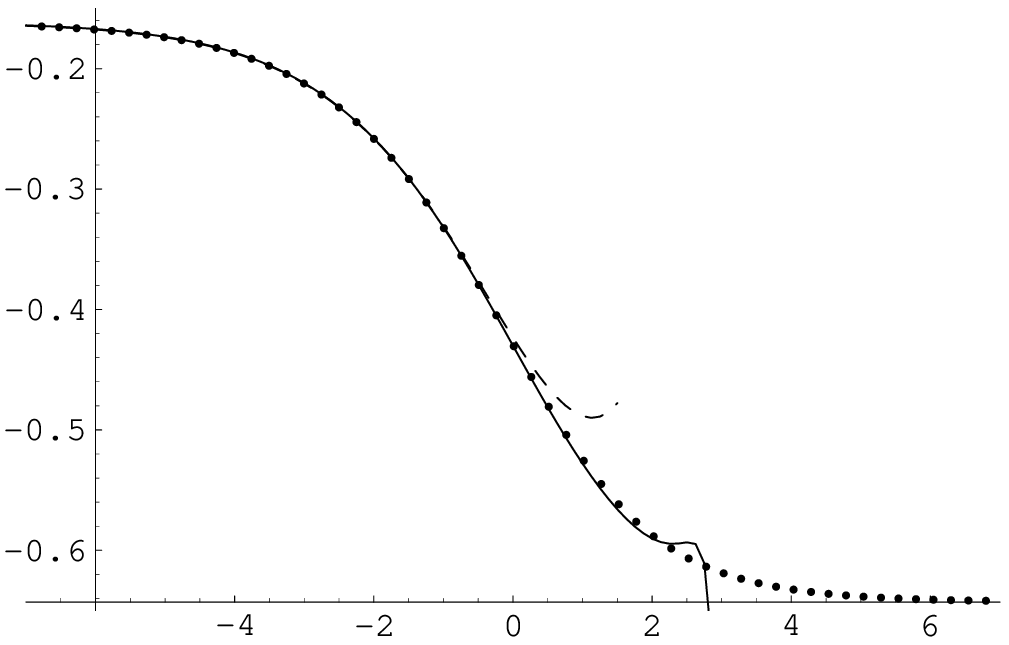} 
\\
\parbox[t]{.75\linewidth}{\small\raggedright%
Figure \ref{fig:Graph16}:
Graphs of $\log( g_\Phi(\tilde h))$ vs.\ $\log(\tilde h)$ 
from the TCSA with 140 states using the exact value of 
$E_0^{\rm circ}$ (dashed line) and the estimated value (solid line)
together with the TBA results (points).
}
\end{array}
\]
}
As a further check, in table \ref{tab:Table3}
we give the TCSA estimates of the coefficients $c_{n0}$. 
{}From \cite{BLZ1}, we know that the TBA and \CPT\ \cgfs\ agree in the
{\em massless} case; in table \ref{tab:Table3} we give the values of 
$c_{00}$, $c_{10}$ and $c_{20}$ from \CPT\ and $c_{30}$ from TBA.
As can be seen, the extrapolations of $c_{00}$
and $c_{10}$ are really quite good but the TCSA under-estimates
the value of $c_{20}$ by about 15\% and the value of
$c_{30}$ by a similar amount. 
However,
the TCSA estimates are still increasing quite fast
with truncation level and would probably agree better on increasing
the truncation level.
{
\small
\refstepcounter{table}
\label{tab:Table3}
\[
%
%
%
%
%
  \begin{array}{c|llllll|l}
      & \multicolumn{6}{c|}{\rm TCSA} 
  
      & \multicolumn{1}{c}{\rm CPT/TBA} \\ 
 
      & \hbox{~67 states}  
      & \hbox{~81 states}  
      & \hbox{~98  states} 
      & \hbox{~117 states} 
      & \hbox{~140 states} 

      & \hbox{~$\infty$ states}

  
      & 
   \\ \hline

 c_{00}  
      &  -0.161720 
      &  -0.161731
      &  -0.161739
      &  -0.161745
      &  -0.161749

      &  -0.16176

      &  -0.1617535656..
  \\

 c_{10}
      & \m  0.997714
      & \m  0.997731
      & \m  0.997742
      & \m  0.997751
      & \m  0.997758

      &  \m 0.99777


      &  \m 0.9977728216..
   \\

 c_{20}  

      &  -0.044147
      &  -0.044428
      &  -0.044674
      &  -0.044890
      &  -0.045083

      &  -0.046


      &  -0.0525515368..
   \\

 c_{30}

    & \m  0.007897
    & \m  0.007965
    & \m  0.008024
    & \m  0.008076
    & \m  0.008122

      &  \m 0.0084

      &  \m 0.00966359^*

   \\
    
  \end{array}
\]
}
\centerline{%
Table \ref{tab:Table3}: 
The coefficients $c_{n0}$ from TCSA, \CPT\, and TBA (denoted by *)
}

\resection{The TBA approach to \gfs}
\label{sec:tba0}

The TBA is an alternative to conformal perturbation theory 
and is characterised by its use of IR scattering data - the bulk
scattering matrix of the massive particles, and the reflection factors
describing the scattering of the particles off the boundaries.

Exact TBA equations describing the
\gf\ flow for both massive and massless bulk were proposed in
\cite{LMSS}, following on from earlier work in \cite{FS1}.
The derivation of these equations (which we recall briefly in section
\ref{sec:tba})
involved a number of assumptions, but to date only
some simple limiting
behaviours have been checked.  We will perform a 
more detailed analysis for the Yang-Lee case, 
which will lead to the conclusion that the equations
of \cite{LMSS} do not tell the whole story in every situation.

\subsection{The Yang--Lee scattering data}

As a massive scattering theory, the Yang--Lee model
can be thought of as a reduction of the Sine-Gordon model 
at coupling constant 
$8\pi/\gamma=3/2$ 
(in the normalisation of \cite{ZZ}).
The bulk  spectrum reduces to a single particle 
(the first Sine-Gordon breather)
with two--particle S--matrix \cite{CM}
\eq
  S(\te)
=  -(1)(2)
\;,\;\;
  (x)
=  {\sinh \lf( \fract{\te}{2} +\fract{i \pi x}{6} \ri)
     \over
    \sinh \lf( \fract{\te}{2} -\fract{i \pi x}{6} \ri)}~.
\en
In \cite{Us1}, the boundary reflection factors
corresponding to the boundary conditions 
$\Phi(h)$ and $\One$ discussed in the previous sections
were identified as
\eq
  R_{\Phi(h)}(\te)
= R_b(\te)
\;,\;\;\;\;
  R_\One(\te)
= R_0(\te)
\;,
\label{eq:ss3}
\en
where 
\eq
  R_b(\te)
= \lf(\fract{1}{2}\ri)\lf(\fract{3}{2}\ri)
  \lf(\fract{4}{2}\ri)^{-1}
  \lf( S(\te+i \pi \fract{b+3}{6})
  S(\te-i \pi \fract{b+3}{6}) \ri)^{-1}
\label{bf}
\label{eq:ss2}
\en
These reflection factors are special cases of the
general Sine-Gordon breather reflection factors proposed in
\cite{Ga} with the parameters $\eta,\vartheta$ of \cite{GZ,Ga} taking the
values 
\[
  \frac{2\eta}\pi
= \pm ( \frac b2 + 2 )
\;,\;\;\;\;
  \frac{2i\vartheta}\pi
= \pm ( \frac b2 + 1 )
\;.
\]
In \cite{Us1} the 
relation between $b$ and $h$ was conjectured on the basis of
numerical fits to be
\eq
  h
= -
  \,
  |h_{c}|
  \,\sin( \pi (b+.5)/5) 
  \,M^{6/5}
\;,\;\;\;\;
  h_{c}
= -0.685289983(9) 
\;.
\label{eq:hnum}
\en
An exact formula for $h_c$ was also proposed, though no derivation
was given; that gap will be filled in section~\ref{sec:gratio}
below.
Taking $b$ real gives $|h|{<}|h_{c}|M^{6/5}$; to obtain
$h{<}{-}|h_c|M^{6/5}$ or $h{>}|h_c|M^{6/5}$ we take 
$b{=}{-}3{-}i\hat b$ and $b{=}2{+}i\hat b$ (with $\hat b$ real)
respectively.

The massless limit is obtained by taking
$M L = l \to 0$ while keeping
$\tilde h $ constant, where
\be
  \tilde h
= h L^{6/5}
= - \, |h_c| \, \sin( \fract \pi 5(b+\fract 12))\, 
    l^{6/5}
\;.
\label{eq:htilde}
\ee
Since $|h/(h_c M)| \to \infty$ in 
this
limit, $b$ takes
the values
$-3-i\hat b$ and $2+i\hat b$
for $\tilde h>0$ and $\tilde h<0$ respectively.

For $\tilde h>0$, 
with $b = -3 - i \hat b$, we have
$  \sin( \fract \pi 5(b+\fract 12))
= - \cosh( {\pi\hat b}/5)
,$
and for $\hat b \gg 0$,
\be
  \tilde h
\sim
  \fract 12 \, |h_c| \, \Big( M L e^{\pi\hat b/6} \Big)^{6/5}
\;.
\label{eq:bhat1}
\ee
Hence the massless limit is 
$l\to 0$, $\hat b \to +\infty$,
$( l e^{\pi\hat b/6} )$ finite.
Substituting $ b = -3 - i \hat b$ into (\ref{eq:ss2}) we find
\be
  R_{-3-i\hat b}(\theta)
= \lf(\fract{1}{2}\ri)\lf(\fract{3}{2}\ri)
  \lf(\fract{4}{2}\ri)^{-1}
  \lf( S(\te + \fract{\pi\hat b}{6})
       S(\te - \fract{\pi\hat b}{6}) \ri)^{-1}
\;,
\ee
which is still a pure phase for real $\theta$. Hence, although $b$ is
now complex, $R_b$ should still describe a physical reflection process.

For $\tilde h<0$,
$b = 2 + i \hat b$, so
$  \sin( \fract \pi 5(b+\fract 12))
= \cosh( {\pi\hat b}/5)$
and for $\hat b \gg 0$,
\be
  \tilde h
\sim
  - \fract 12 \, |h_c| \, \Big( M L e^{\pi\hat b/6} \Big)^{6/5}
\;.
\label{eq:bhat2}
\ee
Hence the massless limit with $\tilde h<0$ is 
$l\to 0$, $\hat b \to +\infty$, $( l e^{\pi\hat b/6} )$ finite.
Substituting $ b = 2 + i \hat b$ into (\ref{eq:ss2}),
\be
  R_{2+i\hat b}(\theta)
= \lf(\fract{1}{2}\ri)\lf(\fract{3}{2}\ri)
  \lf(\fract{4}{2}\ri)^{-1}
  \lf( S(\te - \fract{\pi\hat b}{6} + i \fract{5\pi}6)
       S(\te + \fract{\pi\hat b}{6} - i \fract{5\pi}6) \ri)^{-1}
\;,
\ee
which is not in general a pure phase for real $\theta$. Hence,
$R_b$ should not describe a physical reflection process and (in
particular) we can no longer expect to find a real spectrum in the IR
limit, which was indeed what was found in \cite{Us1} in the TCSA
analysis of the spectrum in the massless case with $\tilde h<0$.

\subsection{The TBA for \gfs}
\label{sec:tba}

The starting-point for the TBA analysis of \cite{FS1,LMSS} was the
 generalisation of the Bethe-ansatz
quantisation equation for particles on a line segment of length $R
\rightarrow \infty$. 
Far from the end points, the scattering is characterised by a two-body  
S-matrix element; information about the reflection of the  particles
at the end points is encoded in the reflection factors 
$R_{\alpha}(\te)$ and
$R_{\beta}(\te)$. 
The main difference with the standard TBA analysis is that the
quantity of interest is a  sub-leading term (order $1/R$ with 
respect to the bulk quantities) of the free energy.     

The initial assumption is that a particle state can be approximated in the
thermodynamic limit ($R \to \infty)$ by a distribution of particles 
in scattering states of given rapidities satisfying self-consistency
equations which simultaneously define a density of possible
particle states 
$\rho(\theta)$ and a density of occupied particle states
$\rho^r(\theta)$, related by
\be
   2 \pi \rho(\theta)
=  
-  2 \pi \etaa \delta(\theta) 
+  2 M R \cosh\theta
+ \phi_\alpha(\theta)
+ \phi_\beta(\theta)
- 2 \phi(2\theta)
+ \int_{-\infty}^\infty \rho^r(\theta') \phi(\theta-\theta') \Dth'
\;,
\label{eq:ppr}
\ee
\[
  \phi(\theta) 
= - i \frac{\D}{\Dth} \log S(\theta)
\;,\;\;\;\;\;\;
  \phi_\alpha(\theta) 
= - i \frac{\D}{\Dth} \log R_\alpha(\theta)
\;.
\]
Assuming that the number of allowed states
with effective particle density $\rho^r(\theta)$ is
$ {\cal N}[\rho^r(\theta)] $,
one can write the partition function as an integral
\be
  Z 
= \sum_{\rm states} e^{-L\,E}
= \int \!\!
  {\cal D}[\rho^r(\theta)]
  \;
  {\cal N}[\rho^r(\theta)]
  \,
  e^{-L\,E[\rho^r(\theta)]}
\;,
\label{eq:pi}
\ee
where the energy of the configuration is
$ 
  E
= \int_{0}^\infty 
  ( M \cosh\theta ) 
  \,\rho^r(\theta)
  \, \Dth
\;.
$
The standard TBA method is to calculate ${\cal N}[\rho^r(\theta)]$
by taking the number of allowed configurations 
in the interval $\Delta\te$ with a fixed number of occupied states
$R\rho^r(\te) \Delta \te$ to be 
\eq
  {(R \rho(\te) \Delta \te )! \over (R \rho^r(\te) \Delta \te)! 
   (R\rho(\te) \Delta \te-R\rho^r(\te)  \Delta \te )!}
\;,
\label{ac}
\en
and replacing the factorials by the two leading terms
$ \log \Gamma(z) \sim z \log z - z + \ldots $
of Stirling's formula, so that the total number of configurations with
a given effective density $\rho^r(\theta)$ is
\be
  {\cal N}[\rho^r(\theta)]
\sim
  \exp\Big[\;
  \int_0^\infty
   \lf( \rho \log \rho -\rho^r \log \rho^r
      -(\rho-\rho^r) \log (\rho-\rho^r) \ri)
  \Dth
  \;\Big]
\;.
\label{eq:ss1}
\en
We can then calculate $Z$ in the limit $R \to \infty$
by the saddle point method, giving the leading behaviour
\[
  \log Z
\sim 
  -R E_0^{\rm circ}(L) 
 + \log( g_\alpha(L)\, g_\beta(L))
\;,
\]
where (extending the range of $\theta$ by symmetry where 
necessary and setting $\epsilon = \log( \rho/\rho^r - 1)\,$)
\be
{
\renewcommand{\arraystretch}{1.4}
\begin{array}{rcl}

   E_0^{\rm circ}(L)
&{\!\!=\!\!}&\ds
   \int_{-\infty}^\infty
   m \cosh\theta\, L(\theta)\,
   \fract{\Dth}{2\pi}
\\[3mm]

  \log( g_\alpha(L)\, g_\beta(L) )
&{\!\!=\!\!}&\ds
   \int_{-\infty}^\infty
   \Big( \phi_\alpha(\te)   + \phi_\beta(\te) - 2\phi(2\theta)
     - 2 \pi \etaa \delta(\te)
   \Big) 
   \, L(\te)\,
   \fract{\Dth}{4\pi}
\label{eq:gf}
\end{array}
}
\ee
where
$L(\theta)= \log( 1 + e^{-\epsilon(\te)})$ and $\epsilon(\theta)$ solves
the equation
\be
  \epsilon(\theta)
=
   mL \cosh\te 
 - \int_{-\infty}^\infty\! 
   \phi(\te - \te')\,
   L(\te')\,
   \fract{\Dth'}{2\pi}
\;.
\label{eq:btba}
\ee
This gives exactly the same result for $E_0^{\rm circ}(L)$
as the usual TBA on a circle \cite{Zb}.
The \gfs\ are then identified as
\be
{
\renewcommand{\arraystretch}{1.4}
\begin{array}{rclrcl}
   \log( g_\alpha(L) )
\;=\;
   \int_{-\infty}^\infty
\Big(\,
   \phi_\alpha(\te) 
 \;-\; 
   \phi(2\te) 
 \,-\, 
   \pi\etaa\delta(\te)
   \,\Big)
   \, L(\te)\,
   \fract{\Dth}{4\pi}
\end{array}
}
\label{eq:gf2}
\ee
There are a number of possible
sources of error in this expression.
Firstly there are possible corrections from the next-to-leading term
$({-}\log z/2)$ in Stirling's formula, and secondly from the 
corrections to the saddle point evaluation of the integral in 
(\ref{eq:pi}).
Finally, the initial assumption that the Hilbert space for the model
on a line segment can be satisfactorily approximated by scattering
states might be wrong.
Unfortunately we do not as yet know how account for these effects in
a consistent way.
In \cite{LMSS} they were formally bypassed by assuming
that they give an overall contribution (also appearing in the
periodic case) which can be  neglected 
when considering changes in the $g$-functions.
Further problems may arise due to the presence of boundary
bound states; again, these were neglected in the above discussion.
In section \ref{sec:ambiguity} below
the corrections that are necessary for the Yang--Lee model in this case
will be found empirically, but a first-principles treatment is still
lacking. With these caveats,  
equations (\ref{eq:btba}) and (\ref{eq:gf2}) form the prediction of
\cite{LMSS}.

\subsection{Properties of the TBA \gfs}

{}From (\ref{eq:btba}) and (\ref{eq:gf2})
it is possible to derive several analytic
properties of the TBA \gfs.

In section \ref{sec:ambiguity} we examine 
possible ambiguities in
the \gfs, and give the prescription which we have adopted in this paper.
In section
\ref{sec:ftba},
under the assumption that
the TBA \gfs\ are the sum of a term linear in
$L$ and a term which is regular in $L^{6/5}$,
the coefficient of the linear term is evaluated analytically.
This enables us to define the TBA \cgfs, which are
directly comparable with the TCSA results.

The behaviour of $L(\theta)$ in the UV and IR limits is well
understood so it is possible to find the net change in the \gfs\ along
the flow.
The \gfs\ in eqn.\ (\ref{eq:gf2}) suffer from ambiguities due to the
possibility to change the integration contour, but
we show in section \ref{sec:netchange} that there are no contours
for which the net change in the TBA \gfs\ agrees with the TCSA result.

In section \ref{sec:gratio} we show that
the expression for the ratio
$\log( g_{\Phi(h(b))}(l) / g_{\One}(l))$ 
is rather simpler than that
for the two \gfs\ separately, as was also the
case in the perturbed conformal field theory analysis.
This enables us to make an expansion in the massless case and obtain
the analytic value of $h_c$ defined in eqn.\ (\ref{eq:hnum}).

We give some numerical power series fits to the TBA results
in section \ref{sec:tbaseries}, and find a combination of functions
which appears to be regular in $l^{12/5}$. 

Finally in section \ref{sec:sings}
we discuss the singularities that occur in $g_{\Phi(h)}$ for
sufficiently negative $\hat h$ and relate these to the singularity in
the ground-state energy on the strip observed in \cite{Us1}.

\subsubsection{The contour prescription for the \gfs}
\label{sec:ambiguity}

If we examine the formula (\ref{eq:gf2}) for the \gf\ in a little more
detail, we notice a couple of immediate problems.
Firstly, the two distinct boundary conditions $\Phi(h(b=0))$ and $\One$
are described by the same reflection factor and hence have
the same kernel function $\phi_0$, yet have different \gfs.
Secondly, $\phi_b$
describing the $\Phi(h(b))$ boundary condition has poles which are
$b$--dependent and (in particular) cross the real axis for $b=\pm 1$.
This means that taking the naive interpretation of (\ref{eq:gf2}),
with the integration contour always along the real axis,
will
lead to \gfs\ which are discontinuous at $b=\pm 1$ whereas the
physical \gfs\ should be continuous in $b$.
The resolution of these two problems appears to be that the
contour in (\ref{eq:gf2}) depends on the boundary
condition, and in particular is different for the $\One$ and
$\Phi(h(b=0))$ boundary conditions. Changing the integration contour
will mean encircling some of the poles of $\phi_b$, the poles so
encircled being called `active', in analogy with the terminology
adopted in \cite{DTa} in a related context.

The exact choice is, at this stage, a matter of trial and error.
In section \ref{sec:netchange} we show that there are no
choices which will give the same \gfs\ as the TCSA, but with the
following prescription at least the ratios of \gfs\ agree: 

\begin{enumerate}

\item{The $\One$ boundary condition}

We take $\log(g_\One)$ to be given by the straightforward application
of (\ref{eq:gf2}), that is 
\[
  \log( g_\One )
=
   \int_{-\infty}^\infty
   \Big(\,
   \phi_0(\te) 
   \,-\, 
   \phi(2\te) 
   \,-\, 
   \pi\etaa\delta(\te)
   \,\Big)
   \, L(\te)\,
   \fract{\Dth}{4\pi}
\;,
\]
with the integration contour along the real axis.

\item{The $\Phi(h(b))$ boundary condition}

We take  
$\log(g_{\Phi(h(b))})$ to be given by (\ref{eq:gf2}) for
$-3\,{<}\,b\,{<}\,{-}1$, that is  
\[
  \log( g_{\Phi(h(b))} )
=
   \int_{-\infty}^\infty
   \Big(\,
   \phi_b(\te) 
   \,-\, 
   \phi(2\te) 
   \,-\, 
   \pi\etaa\delta(\te)
   \,\Big)
   \, L(\te)\,
   \fract{\Dth}{4\pi}
\;,\;\;\;\; -3 < b < -1\;,
\]
with the integration along the real axis.
For all other values of $b$, it is given by the
analytic continuation of
this function.
In particular, for $\hat h{>}|h_c|$ (i.e. $b{=}{-}3{-}i\hat b$)
in the massive case, and for $\tilde h>0$ in the massless, the
integration contour may again be taken along the real axis.

\end{enumerate}

\noindent
The simplest way to derive the continuation of $\log(g_{\Phi(h(b))})$
is to isolate the terms in the integrand which have $b$--dependent
poles and treat these separately.
We have
\[
  \phi_b(\te) 
= \phi_0(\te)
 - \phi(\te - i\pi\fract{b+3}6)
 - \phi(\te + i\pi\fract{b+3}6)
\;,
\]
so that for $-3\,{<}\,b\,{<}{-}1$,
\be
  \log( g_{\Phi(h(b))} )
= \log( g_\One )
- \int_{-\infty}^{\infty} 
  \Big(
  \phi(\te - i\pi\fract{b+3}6)
+ \phi(\te + i\pi\fract{b+3}6)
  \Big)
  L(\te)\,
  \fract{\D\te}{4\pi}
\;.
\label{eq:I1}
\ee
When the integration contour is along the real axis (but not
necessarily otherwise) the two terms in the integral (\ref{eq:I1})
give the same contribution, so at least
for $-3\,{<}\,b\,{<}{-}1$ we have
\bea
  \log( g_{\Phi(h(b))}) - \log(g_\One)
&=& 
- \int_{-\infty}^{\infty} 
  \phi(\te - i\pi\fract{b+3}6)\,
  L(\te)\,
  \fract{\D\te}{2\pi}
\label{eq:I2}
\eea
The term in the integral
$  \phi(\te - i\pi\fract{b+3}6)  $
has poles at 
\be
  \te 
= i\pi\fract{b-5}6 + 2n\pi i
\;,\;\;\;\;
  \te 
= i\pi\fract{b-1}6 + 2n\pi i
\;,\;\;\;\;
  \te 
= i\pi\fract{b+1}6 + 2n\pi i
\;,\;\;\;\;
  \te 
= i\pi\fract{b+5}6 + 2n\pi i
\;.
\label{eq:poles}
\ee
This means that as $\te$ passes from $b<-1$ to $b>-1$, the contour
must deform away from the real axis to avoid a discontinuity from the
pole at $\te = i\pi(b{+}1)/6$, or equivalently the pole at $i\pi(b{+}1)/6$
is `active' for $b>-1$,
and similarly the pole at $i\pi(b{-}1)/6$ is `active' for $b>1$,
as shown in figure \ref{fig:contour1}.

Note that the poles that are `active' are exactly those
associated with boundary bound states \cite{Us1,Us1b}.
It is clear that the transition of a scattering state into a boundary
bound state is always associated with a pole in $\phi_\alpha$ crossing
the real axis, and in this case the converse is also true. 
(Recall that
the general interpretation of simple poles can be subtle, and not
all correspond to boundary bound states -- see \cite{Us1b}.)
Hence, 
this seems the natural prescription for the integration contour in the
presence of such bound states and it would be interesting to arrive at
this result from first principles and check it in further models.

\newcommand{\boxed}[1]{{\setlength{\fboxsep}{4mm}\fbox{ $\ds{#1}$}}}
\[
\begin{array}{c}
\refstepcounter{figure}
\label{fig:contour1}
\epsfxsize=.8\linewidth
\boxed{\epsfbox{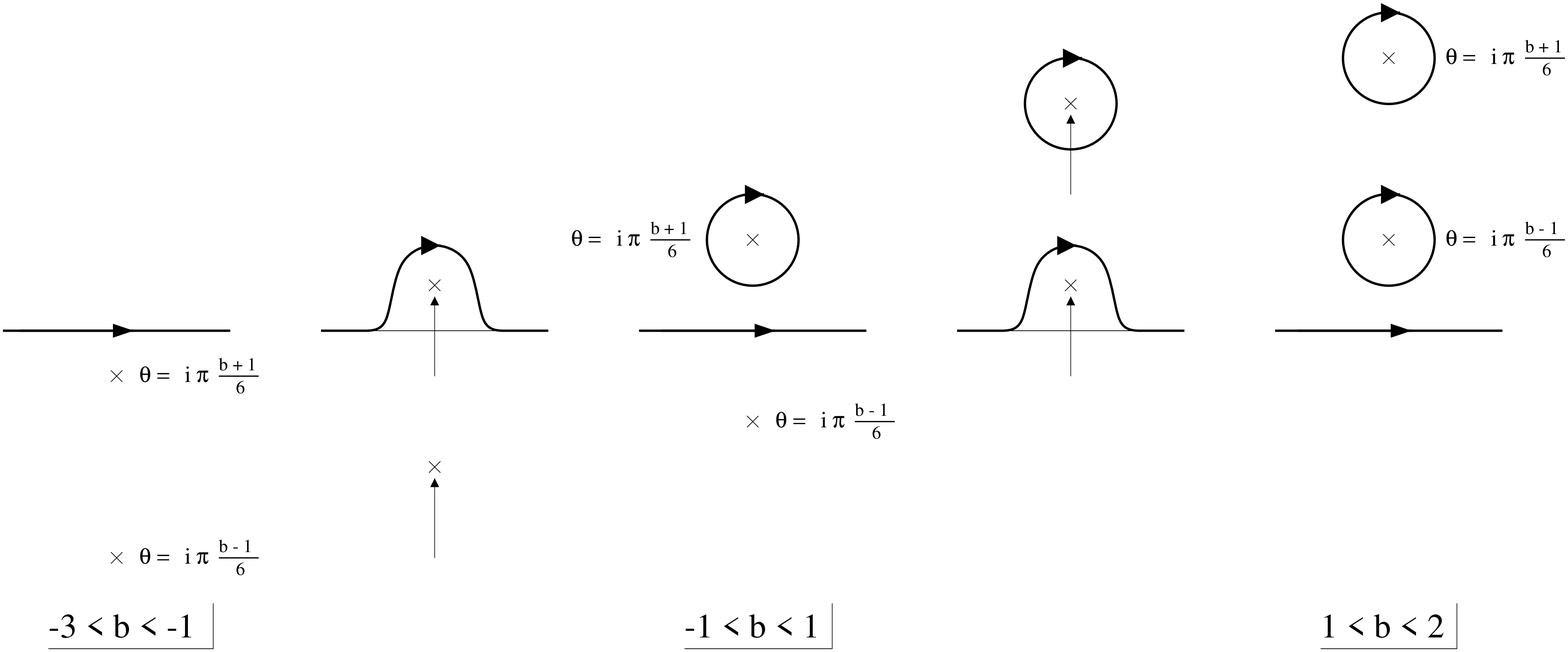}}
\\
\parbox[t]{.7\linewidth}{\small\raggedright%
Figure \ref{fig:contour1}:
The determination of the contour through the analytic continuation of
$\log(g_{\Phi(h(b))})$ from $b{<}{-1}$ to $b{>}1$. 
}
\end{array}
\]

\subsubsection{Analytic structure of the TBA \gfs.}
\label{sec:ftba}

We expect the \gfs\ as evaluated in the TBA to be regular functions of
$L^{6/5}$ after the removal of an `extensive' part, linear in $L$. 
Thus they should have the form
\[
  \log g_\alpha 
=  L f_\alpha + \log \cg_\alpha
\;.
\]
Under this assumption, 
the coefficient of the linear term can be evaluated analytically
as
\[
  f_\alpha
= - \,M\,\frac{ \D}{\D l} \, g_\alpha(l)\, \Big|_{l=0}
\;.
\]
Al.~Zamolodchikov showed in \cite{Zb} that,
in the limit $l \to 0$,
  $\te > 0$,
\[
  L(\theta)
\sim
  L_{\rm kink}(\theta - \log\fract 2l)
\]  
with
\[
  \int_{-\infty}^\infty
   e^{-\theta}
 \, \frac{\D L_{\rm kink}(\theta)}{\D \theta}
 \, \D\te
= - \frac \pi{\sqrt 3}
\;.
\]
Substituting this into (\ref{eq:btba}) we arrive at
\be
  {f^{\rm TBA}_b}
= \left( \fract{\sqrt 3 - 1}4 + \sin\fract{\pi b}6 \right)\, M
\;,
\label{eq:ftba}
\ee
independent of the contour prescription we choose.
In the massless limit, using
(\ref{eq:bhat1}) we find for $\tilde h>0$ that 
\be
   f_b^{\rm TBA} L
\sim
  -\, 2^{-1/6}
  \left|\,
  { \tilde h }/{ h_c }
  \,\right|^{5/6} 
\;.
\label{eq:fb2}
\ee

\subsubsection{Net change in the TBA \gfs\ along the flows}
\label{sec:netchange}

Given (\ref{eq:gf2}), it is straightforward to calculate 
$\log( g_{\alpha} )$ in the two limits
$L \rightarrow 0$ and $L \rightarrow \infty$ using the result
\be
  L(\te) \to
\cases{
  \log(\fract {1+\sqrt 5}2 )
\;,\;\;\;\; & UV, \cr
  0
\;,\;\;\;\; & IR.
}
\label{eq:Llim}
\ee
While the IR result is simple, 
\be
  \log(\, g_\alpha\,) \Big|_{\rm IR}
= 0
\;,
\label{eq:lggIR}
\ee
the UV limit depends on the choice of integration contour in
(\ref{eq:gf2}). Although we have given above a particular
prescription for the contours, no matter which contour is chosen, one
{\em must} end up with
\be
  \log(\, g_\alpha\, ) \Big|_{\rm UV}
= \fract n4\, 
  \log(\fract { 1 + \sqrt 5}2 )
\;,\;\;\;\;
  n \in \mathbb{Z}
\;.
\label{eq:lggUV}
\ee
Therefore the most general prediction of (\ref{eq:gf}) would be 
an interpolating flow
\be
  \log( g_\alpha )\Big|_{\rm UV}
\;-\;
  \log( g_\alpha )\Big|_{\rm IR}
= \fract n4\, 
  \log(\fract {1 +  \sqrt 5} 2 )
\;,\;\;\;\;
  n \in \mathbb{Z}
\;.
\label{eq:lggflow}
\ee
There are two distinct physical situations: massless bulk and massive
bulk.

\begin{enumerate}

\item{Massless bulk}\\
There is a single identifiable massless flow with $\tilde h>0$, from
$\Phi$ to $\One$, for which 
(\ref{eq:gvalues}) gives
\be
  \log( g_\Phi)
\;-\;
  \log( g_\One)
=
\log( \fract {1+\sqrt 5}2 )
\;,
\label{eq:lggflow1.5}
\ee
and which is in agreement with (\ref{eq:lggflow}) for $n=4$.

\item{Massive bulk}\\
For the massive flows from $\One$ and from $\Phi(h)$ with
$\hat h{>}{-}|h_c|$, $g_{UV}$ are given by conformal field theory, 
and our expectation that for the massive theory $g|_{IR}=1$ 
(already proposed in \cite{Us1} and well-supported by the TCSA
results) leads to the results
\bea
  \log( g_\One)\Big|_{\rm UV}
\;-\;
  \log( g )\Big|_{\rm IR}
&=&
  - \fract 14 \log\left| \fract{1 + \sqrt 5}2 \right|
  - \fract 18 \log 5
\;,
\nn\\
  \log( g_{\Phi(h)})\Big|_{\rm UV}
\;-\;
  \log( g )\Big|_{\rm IR}
&=&
  \m \fract 34 \log\left| \fract{1 + \sqrt 5}2 \right|
  - \fract 18 \log 5
\;.
\label{eq:lggflow2}
\eea
These are in contradiction with (\ref{eq:lggflow}).

\end{enumerate}

\noindent
Thus we see that 
the TBA equations (\ref{eq:btba}) do not
agree with the TCSA even at the level of the overall change in the
\gfs\ without investigating finer detail.

We note however that the ratio $\log( g_\Phi(h)/g_\One)$ also changes by
$\log( \fract 12(\sqrt 5+1))$ along a massive flow, so the TBA
equations do have the possibility to describe the ratio correctly, and
indeed we shall find supporting evidence for this later on.

\subsubsection{The relation of the TBA \gfs\ to the Y--function}

Having written the ratio of the \gfs\ in the form (\ref{eq:I2}), 
we observe that the integral on the right hand side is
exactly of the form appearing in the TBA equation determining the
pseudo-energy $\epsilon$:
\bea
  \log( g_{\Phi(h(b))}) - \log(g_\One)
&=& 
- \int_{-\infty}^{\infty} 
  \phi(\te - i\pi\fract{b+3}6)\,
  L(\te)\,
  \fract{\D\te}{2\pi}
\nn\\
&=& 
  \epsilon(i\pi\fract{b+3}6)  
+ l\sin(\fract{b\pi}6)
\;.
\label{eq:I2b}
\eea
If we absorb the linear term into the \gfs\ following (\ref{eq:ftba})
we obtain exactly the \cgfs, so that 
the \gfs\ which follow from the TBA equations of \cite{LMSS} 
predict the particularly simple form
for the ratio of the \cgfs
\be
  \log
  \left(\,
  { \cg_{\Phi(h(b))} }\,/\,{\cg_\One }
  \,\right)
= \epsilon( i \pi \fract{ b+3}6 )
=  \log Y(i\pi\fract{b+3}6 )
\;.
\label{eq:cgratio}
\ee
The same result  was also obtained in \cite{BLZ1} by a different
method, but with the restriction to the massless case. Given this, the
extension (\ref{eq:cgratio}) to the massive case is perhaps not so
surprising, but the derivation via the TBA seems unrelated to the
methods of \cite{BLZ1}.
Just to finish this section, we comment that we have subjected
(\ref{eq:cgratio}) to several tests outlined in later sections,
all of which support this identity.

\subsubsection{The expansion of $\log(\,g_{\Phi(h(b))}(l)\,)$
in the massless limit}
\label{sec:gratio}

The perturbative expansion of $\epsilon(\theta)$ in the massless limit
has been found in \cite{BLZ1,BLZ2}.
If $\epsilon(\theta)$ is normalised so that for large 
$(l \exp \theta)$ it has the asymptotic behaviour
$ \epsilon(\theta) \sim \frac {l\,  e^\theta}2 $,
then for small $(l \exp \theta )$ it has an expansion
\[
  \epsilon(\theta)
= \log \frac{1 + \sqrt 5}2
\;+\; 
   C_1\,(l\,e^\te)^{6/5}
\;+\; 
   C_2\,(l\,e^\te)^{12/5}
\;+\;
   \ldots
\;,
\]
where (from appendix A of \cite{BLZ2})
\[
  C_1 
= \frac
  { 2^{-7/5} \pi^{7/5} \Ga( \fract 15) }
  { \cos \fract \pi 5  \Ga( \fract 23) \Ga(\fract 35)^2 \Ga(\fract 45)}
  \,
  \left( \frac { \Ga( \fract 23 ) }{ \Ga( \fract 16 ) } \right)^{6/5}
\;,\;\;\;\;
  C_2 
= \left( \fract 1{\sqrt 5} - \fract 12 \right)\,
  C_1^2
\;.
\]
{}From perturbed conformal field theory 
the coefficients in
the expansion (\ref{eq:cddefs}) are
\be
  \log \frac{\cg_{\Phi(h)}(0,L)}{\cg_{\One}(0,L)}
= \log \frac{g_\Phi}{ g_\One}
 \;+\;  c_{10}\, (h L^{6/5})
 \;+\;  c_{20}\, (h L^{6/5})^2
 \;+\; \ldots
\ee
and were found
in section \ref{sec:boundaryp}
to be 
given by (\ref{eq:c10c20}):
\be
  c_{10} 
= 
  (2\pi)^{-1/5} 5^{1/4}
  \left|
  \frac{ \Gamma(\fract 25) \Gamma( \fract 65 ) }
       { 2 \cos(\pi/5) \Gamma(\fract 45)^2     }
  \right|^{1/2}
\mathematica{0.997772821597 
 c1pcft = (2 Pi)^(-1/5) 5^(1/4) * 
          Sqrt[ Gamma[2/5] Gamma[6/5] / 
          ( 2 Cos[Pi/5] Gamma[4/5]^2 ) ]
}
\;,\;\;\;\;
  c_{20} 
=
  \left( \fract1{\sqrt 5} {-}\fract 12 \right)
  (c_{10})^2
\;.
\ee
Given that in the massless limit we have 
\be
  \log \frac{\cg_{\Phi(h)}(0,L)}{\cg_{\One}(0,L)}
= \epsilon( \frac{\pi \hat b}6 )
\;,\;\;\;\;
   h L^{6/5} 
= 
  - \fract 12
  \, h_c
  \, (l e^{\pi\hat b/6})^{6/5}
\;,
\label{eq:gmassless}
\ee
we deduce that
\be
  h_c
= - \frac{ 2 C_1 }{c_{10}}
= - \pi^{3/5}
  \, 2^{4/5}
  \, 5^{1/4}
  \frac{ \sin \fract{2\pi} 5 }
       { ( \Ga( \fract 35 ) \Ga(\fract 45) )^{1/2} }
  \left( 
  \frac{ \Ga( \fract 23 ) }{\Ga( \fract 16 )}
  \right)^{6/5}
\mathematica{
 - Pi^(3/5) * 
  2^(4/5) *
  5^(1/4) *
  (   Sin[ ( 2 Pi) / 5 ]  / 
      ( Gamma[3/5] Gamma[4/5] )^(1/2) ) * 
  ( Gamma[2/3] / Gamma[1/6] )^(6/5)
}
%
%
= -0.68528998399118\ldots
\;.
\label{eq:hcexact}
\ee
This is in perfect agreement with the numerical result obtained in
\cite{Us1}. 
  
\vskip 5mm

\subsubsection{Series expansions of ingredients of the TBA expressions}
\label{sec:tbaseries}

In an attempt to understand better the discrepancies between the TBA
results and those from other approaches, 
we have investigated numerically the various ingredients of the TBA
\gfs.
The aim was to determine their analytic structure by various
power series fits in the variables
$x\equiv \lambda L^{12/5}$ and $y\equiv h L^{6/5}$.
These results give highly accurate determinations of the first few
coefficients, but the accuracy rapidly tails off.

In each case we initially assumed an analytic expansion in $x^{1/2}$
with
  the addition of possible non-analytic terms of the form
$x^{5n/12}$, with the results given in table \ref{tab:Table4}.
On the basis of these results we then performed a fit to more
restricted functional forms (with fewer non-analytic terms) with the
results as in table \ref{tab:Table5}.
Finally, we know on theoretical grounds that $\epsilon(i\pi(b+3)/6)$
has an expansion in $x$ and $h$ and we give the numerical
determination of these coefficients in table \ref{tab:Table6}.
{\small
\[
\refstepcounter{table}
\label{tab:Table4}
\renewcommand{\arraystretch}{1.4}
\begin{array}{|l|l|}

\hline

&
- 0.240605912529803  
- 2.2~10^{-14}                   \, x^{5/12}
+  0.419185703387               \, x^{1/2}
\\   

- \fract{1}{2} L(0)\, \equiv \fract{1}{2} \epsilon(0) -\epsilon(i \pi/3)

\m

&
- 3.~10^{-12}                   \, x^{10/12}   
- 0.26930751                    \, x 
- 3.6~10^{-8}                    \, x^{15/12}
+  0.0120551        \, x^{3/2}

\\

&
-  1.~10^{-7}       \, x^{20/12}
+  0.07138          \, x^{2}
-  2.5~10^{-5}       \, x^{25/12}
-  0.0096           \, x^{5/2} 

+ \ldots

\\\hline

&
0.481211825059603
-  4.~10^{-14}                   \, x^{5/12}
-  0.67825671567                 \, x^{1/2}

\\   

\m L(\fract{i \pi}6)\, 
\m

&
-  9.1~10^{-13}                   \, x^{10/12}
+  0.308598962                   \, x 
-  7.~10^{-8}                    \, x^{15/12}
+  0.05575394                   \, x^{3/2}

\\

&
-  1.3~10^{-7}                   \, x^{20/12}
-  0.08108                      \, x^{2}
-  2.6~10^{-5}                   \, x^{25/12}
-  0.023                        \, x^{5/2} 
+ \ldots

\\\hline

&
\m 0.240605912529
- 7.3~10^{-10}         \,x^{5/12}
- 0.67825671          \, x^{1/2}

\\

- \inti  \phi(2 \te)\, L(\te) \, \fract{\D\te}{2\pi}
\m

&
+ 0.8102843           \, x^{10/12}
- 0.33287            \, x
- 0.0002             \, x^{15/12}
\\    
&
+ 0.062              \, x^{3/2}
- 0.045              \, x^{20/12}
 +\ldots

\\\hline

&
- 0.4812118250596           
+ 0.967384888      \,x^{5/12}
- 4.3~10^{-9}      \,x^{1/2}

\\ 

\m \inti   
   \phi_0(\te)\,L(\te)\,
   \frac{\D\te}{2\pi}

&
-  0.8102843         \,x^{10/12} 
+  0.20572          x
+  0.00026           x^{15/12}
\\
&
+

\ldots

\\\hline

&
-  0.2406059125298
+  0.96738488748    \, x^{5/12} 
-  0.6782567157         \, x^{1/2}
\m

\\ 

   \inti   
   (\phi_0(\te) - \phi(2\theta))
   \,L(\te)\,
   \frac{\D\te}{2\pi}

&
+ 6.~10^{-9}      \, x^{10/12}
- 0.1271498       \, x
+ 5.~10^{-7}      \, x^{15/12}
+ 0.055748         \, x^{3/2}
\\
&

+ 9.~10^{-6}       \, x^{20/12}
-0.02350           \,x^2
+0.00004           \, x^{25/12}
+\ldots
\\\hline

\multicolumn{2}{c}{
\hbox{
Table \ref{tab:Table4}:
the results of inital fits to the TBA data
}}
\end{array}
\]

\[
\refstepcounter{table}
\label{tab:Table5}
\renewcommand{\arraystretch}{1.4}
\begin{array}{|l|l|}

\hline

&
- 0.240605912529802 
+  0.4191857033888  \,x^{1/2}
- 0.269307520 \,x 
\\   

- \fract{1}{2} L(0)
\m

&
+  0.0120552       \,x^{3/2}
+  0.07138         \,x^{2}
-  0.0096        \,x^{5/2} 

\\

&
+ \ldots

\\\hline

&
0.481211825059603
-  0.678256715680  \,x^{1/2}
+  0.30859896278\,x 
\\   

\m L(\fract{i \pi}6)\, 
\m

&
+  0.055753953      \,x^{3/2}
-  0.081082         \,x^{2}
-  0.023            \,x^{5/2} 

\\

&
+ \ldots

\\\hline

&
\m 0.240605912529805
-  0.67825671              \, x^{1/2}
+  0.8102848              \, x^{10/12}

\\

- \inti  \phi(2 \te)\, L(\te) \, \fract{\D\te}{2\pi}
\m

&
-  0.332875       \, x
-  0.00020      \, x^{15/12}
+  0.0615         \, x^{3/2}
\\    
&
-  0.043    \, x^{20/12}
+ \dots
\\\hline

&
-   0.481211825059593          
+   0.96738488736        \,x^{5/12}
-   0.8102849            \,x^{10/12} 

\\ 

\m \inti   
   \phi_0(\te)\,L(\te)\,
   \frac{\D\te}{2\pi}

&
+  0.20572                \,   x
+  0.00021                  \,  x^{15/12}
-  0.0060                   \,  x^{3/2}
+  0.043                    \,  x^{20/12} 

\\
&
+  \dots

\\\hline

&
- 0.240605912529802
+ 0.9673848874698    \, x^{5/12}

\m

\\ 

   \inti   
   (\phi_0(\te) - \phi(2\theta))
   \,L(\te)\,
   \frac{\D\te}{2\pi}

&
- 0.6782567156814      \, x^{1/2}
- 0.127149765        \, x
+ 0.05575395          \, x^{3/2}

\\
&
+ 0.023491 \, x^2
-0.020  \, x^{5/2}
+\ldots

\\\hline

\multicolumn{2}{c}{
\hbox{
Table \ref{tab:Table5}:
the results of more restricted fits to the TBA data
}}
\end{array}
\]
}

{\small
\refstepcounter{table}
\label{tab:Table6}
\footnotesize
\[
\renewcommand{\arraystretch}{1.4}
\begin{array}{|l|}
\hline

\multicolumn{1}{|c|}{
\vphantom{\Big|}
    \ep(\imath \pi \fract{b+3}{6})
 =   m L \sin \fract{\pi b}{6} -
    \inti \phi( i \pi \fract{b+3}{6} -\te) L(\te)  
     \,\D\te/(2\pi)
}

\\
\hline

\m 0.48121182508 
 + 0.332882394  \,x     
 - 0.09242282   \,x^2 
 + 0.040463     \,x^3  
 - 0.02047 \,x^4 
 + 0.0104  \,x^5
 - 0.00385 \,x^6 
\\
 + 0.9977728224  \, y  
 - 0.086482609 \,x y     
 + 0.03812851 \,x^2 y
 - 0.01991664 \,x^3 y
 + 0.010374   \,x^4 y 
 - 0.00377    \,x^5 y
\\ 
 - 0.052551643 \,y^2
 + 0.02553410  \,x y^2   
 - 0.0148490   \,x^2 y^2 
 + 0.008372    \,x^3 y^2
 - 0.003206    \,x^4 y^2
\\
 + 0.00966359 \,y^3
 - 0.0076724  \,x y^3 
 + 0.005150   \,x^2 y^3 
 - 0.00225    \,x^3 y^3
\\
 - 0.0021742  \,y^4
 + 0.002199   \,x y^4 
 - 0.00119    \,x^2 y^4
\\
 + 0.000563   \,y^5
 - 0.00046    \,x y^5
 - 0.00016    \,y^6
\\
\hline
\multicolumn{1}{c}{
\hbox{Table \ref{tab:Table6}:
numerical fit to the expansion of $\epsilon$}
}
\end{array}
\]
}

We immediately notice that with the contour  prescription of section
~\ref{sec:ambiguity}, 
not only does  $g_\One$ not have the correct net overall change, but
neither is $\cg_\One$  a pure power series in $x$ ($\lambda$).
Even sacrificing universality, with the lack of such a  basic
property  there is no hope to identify this  quantity  with  a
properly defined \gf. 

However, we observe that it is possible to find 
a function which has a regular expansion in $x$ from the various
ingredients of the TBA. If we drop the term $-L(0)/2$ 
(which was introduced in \cite{LMSS} to account for the absence of
a zero-momentum single particle state on the strip), and further, instead
of running the contour on the real axis, we instead also 
encircle the singularity of  $\phi_0$ at $\theta= i \pi /6$ clockwise
to pick up an extra contribution 
$-L(i \pi/6) \equiv -\epsilon(i \pi/2)$, we arrive at the function
\[
   \inti   
   (\phi_0(\te) - \phi(2\theta))
   \,L(\te)\,
   \frac{\D\te}{2\pi}
- \epsilon( \fract{i\pi}2 )
\;,
\]
for which we find the expansion
\[
\begin{array}{l}
- 0.7218177375894 
+ 0.967384887468          \, x^{5/12}
+ 2.~10^{-13}         \, x^{1/2}
- 0.435748727             \, x
\\
\m+ 6.~10^{-10}            \, x^{3/2}
+ 0.10457              \, x^{2}
+ 0.004              \, x^{5/2}
\end{array}
\]
which has (within our accuracy)  better prospects of 
being a purely analytic function of
$x$, plus a term in $x^{5/12}$ which we can identify as 
$ (\sqrt 3 - 1) \kappa \, x^{5/12} / 2
= (\sqrt 3 - 1)  l / 2$.

To conclude, the \gfs\ predicted by the TBA equations of \cite{LMSS}
do not  match the quantities derived  perturbatively 
in this paper, nor do they have the expected  analytic
properties. Although  we have partial evidence that the  latter
problem could be overcome with an ad-hoc contour prescription, at the
present stage of our understanding we have no definite justification
for the above-mentioned discrepancies.

\subsubsection{
Singularities for ${\hat h<-|h_c|}$
}
\label{sec:sings}

In  section~\ref{sec:ambiguity}, we made the following
identification 
in the case of massive bulk:
\eq
  \log( g_{\Phi(h(b))}) - \log(g_\One)
= 
  \epsilon(i\pi\fract{b+3}6)  
+ l\sin(\fract{b\pi}6)
\;.
\label{eq:I21}
\en
The TCSA analysis in \cite{Us1} revealed
 the existence of a critical value of $hM^{-6/5}$, in the range
$-0{\cdot}8$ to $-0{\cdot}6$, for which $\log( g_{\Phi(h(b))}g_\One)$
diverges --- this was signalled by a pole in a rational fit to the
\gf. 
Furthermore, the ground state energy on the strip 
has a square--root singularity
at some real value  of  $r$ for  
$h  M^{-6/5} <  h_{c} M^{-6/5}= -0.68529..$
(see section~6 of  \cite{Us1}).
This led us to the  conjecture that the singularities in the two
quantities  occur at the same critical threshold $ h_{c} M^{-6/5}$.
Armed with  relation~(\ref{eq:I21}) we can now prove this.
First note that the quantity    
$Y(\theta,r)=\exp({\epsilon(\theta,r)})$  
is entire in $\theta$; further, for $r\equiv RM$ real, it has zeroes
on the line
${\rm Im\,}\theta= 5 \pi/6$ (or equivalently ${\rm Im}\,\te = -5 \pi/6$),
arranged symmetrically about the imaginary axis.
While their exact locations $\te_n$ are not known, for large $r$ a
good approximation is 
$\sinh({\rm Re}\, \theta_n ) = \pm   \pi (2 n+1) /r$
($n=0,1,2,\dots)$.
Hence as $r$ grows these zeroes move towards 
${\rm Re\,} \theta =0$.

Suppose now that ${\rm Re}\, b=2$ so that the value
$\theta(b)=i\pi\fract{b+3}{6}$ 
relevant for (\ref{eq:I21})
lies on the line ${\rm Im}\,\te=5\pi/6$.
As $r$ increases,
the zeroes $\te_n$ of $Y$
move along the line ${\rm Im}\,\te=5\pi/6$ and at some critical value
$r_c(b)$ one will pass through $\te(b)$, so that
$Y(\theta(b),r_c(b))=0$.
This confirms the presence of logarithmic 
singularities in~(\ref{eq:I21}), fixes the  critical  threshold at
$b{=}2$ and  proves (through eq. (\ref{eq:hnum}))  
the above mentioned conjecture.
A similar analysis shows that logarithmic singularities appear
in the massless $h<0$ case too.

\resection{Comparison of TBA, \CPT\ and TCSA results}
\label{sec:compare}
\label{compa}

Firstly, it is 
a consequence of
\cite{BLZ1} that the TBA and \CPT\ methods
give the same answer for the \cgf\ along the massless flow.
This is borne out by the comparison of our TCSA data with the TBA. 
However, we find disagreement between the TCSA and TBA predictions for
the massive flows.

On general grounds we showed in section \ref{sec:gpcft} that the
perturbative parts of the \gfs, denoted by \cgfs, should have a series
expansion for small $L$ in $L^{6/5}$. 
However, when we try to combine the power series expansions for the
various ingredients given in section \ref{sec:tbaseries} we see that
the \cgfs\ predicted by the massive TBA equations of \cite{LMSS} do
not have the correct behaviour. This is compounded by the fact that in
the massive case the net change in the \gfs\ does not agree with the
CFT predictions. 
This leads us to our conclusion that the \gfs\ and \cgfs\ given by the
massive TBA equations of \cite{LMSS} do not agree with those obtained
viewing the theory as a perturbed conformal field theory.

One of the most natural sources of this discrepancy is from a
neglected sub-leading term in the saddle point evaluation of $\log Z$,
but we have not been able to calculate such a term. 
What is clear is that the discrepancy between the TBA and \CPT\
calculations of the \gfs\ is independent of the boundary conditions.
This is consistent with our observations
in tables \ref{tab:Table2b} and \ref{tab:Table3} that 
the first few coefficients in the power series expansions obtained
from the TBA and TCSA for the {\em ratio} of \cgfs,
$\log(\cg_{\Phi(h)} / \cg_\One)$
do agree and that the extensive free-energy terms (the linear
behaviour of the \cgfs\ for large $L$) are also the same in the TBA
and TCSA results.

\refstepcounter{table}
\label{tab:Table2b}
\[
  \begin{array}{c|l|l|l}
      & \multicolumn{1}{c|}{\hbox{TCSA ($\infty$ states)} }
      & \multicolumn{1}{c|}{\rm TBA} 
      & \multicolumn{1}{c}{\rm exact} 
 \\
\hline
 

 c_{01} - d_1 
      &  \m\m 0.332

      &  \m 0.332882394

      &  \m 0.332882407..
   \\

 c_{02} - d_2
      &  \m{-}0.092

      &  -0.09242282

      &  \multicolumn{1}{c}{\hbox{---}}
   \\

 c_{03} - d_3
      &  \m\m 0.039

      &  \m 0.040463

      &  \multicolumn{1}{c}{\hbox{---}}
   \\

 c_{03} - d_3
      &  \m{-}0.02

      &  -0.02047

      &  \multicolumn{1}{c}{\hbox{---}}
   \\

  \end{array}
\]
\centerline{%
Table \ref{tab:Table2b}:
The coefficients $(c_{0n}{-}d_n)$ from TCSA, TBA and \CPT
}
\\

As further evidence we show in figure \ref{fig:Graph17} the TCSA and
TBA estimates for  the ratio of 
$g_{b=-1.1} $
and 
$g_{b=-1.2} $.

\[
\begin{array}{c}
\refstepcounter{figure}
\label{fig:Graph17}
\epsfxsize=.45\linewidth
\epsfbox[0 50 288 238]{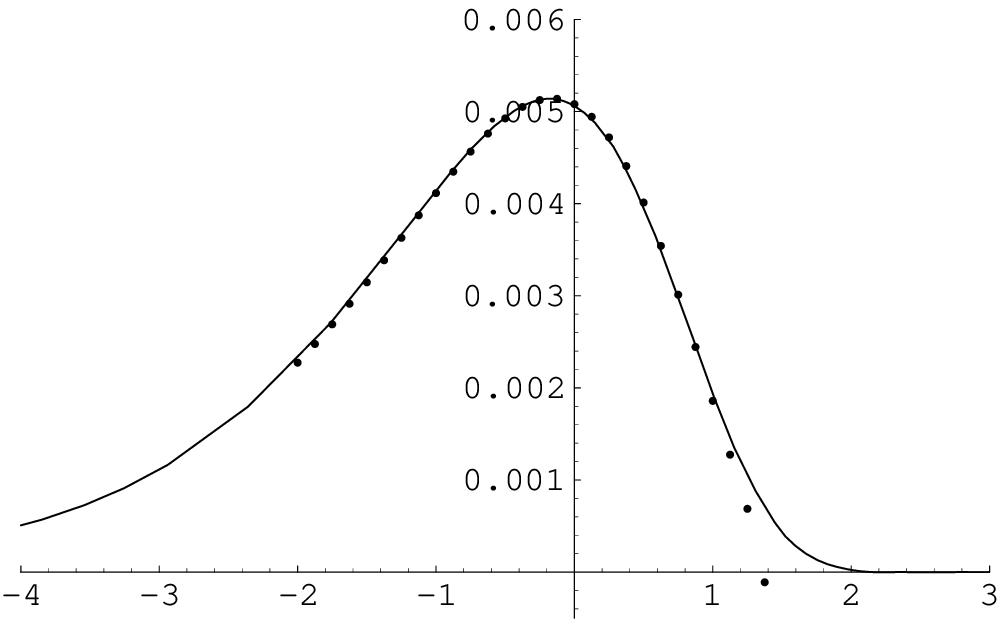}
\\
\parbox[t]{.55\linewidth}{\small\raggedright%
Figure \ref{fig:Graph17}:
The difference
$\log(g_{b=-1.1}) - \log(g_{b=-1.2})$ vs.\ $\log l$
for the massive YL model.
The solid line is the TBA result and the points are from the TCSA to
117 states.
}
\end{array}
\]


\resection{Conclusions}

We have investigated the relation between the \gfs\ proposed on the
basis of a TBA analysis in \cite{LMSS} and those obtained from
perturbed conformal field theory.

We found that the massless case is correctly described by the TBA \gfs\
(up to an overall constant) but that the massive case is not.
However, in all cases, the ratios of \cgfs\ for different boundary
conditions of the same model are correctly predicted by the TBA.
A consequence of the TBA proposal of \cite{LMSS} is that this ratio of
perturbative \cgfs\ should satisfy
\be
  \frac{\cg_{\Phi(h(b))}}{ \cg_\One}
= Y(i\pi\fract{b+3}6)  
\;.
\label{eq:I2c}
\ee
This relation is already contained in the results of \cite{BLZ1}
in the {\em massless} case, but the derivation there seems rather
different to ours.
We have tested (\ref{eq:I2c}) in the massive case in two ways:
we have compared the power series expansions from fits to the TBA data
to the perturbed conformal field theory and TCSA calculations of these
expressions; we have further compared the ratio of \cgfs\ directly
over a range of $l$ using the TBA and TCSA.
Both tests support (\ref{eq:I2c}).

We have also found strong support from the TCSA for the correctness of
the TBA calculation of the `extensive' boundary free energy
$f_\alpha$.

To summarise, we have found strong evidence for the relation
\[
  g^{\rm CFT}_\alpha(l)
= g^{\rm TBA}_\alpha(l)
+ k(l)
\;,
\]
where $k(l)$ is an (unknown) function of $l$ vanishing
at $l=\infty$, and which is independent of the boundary
conditions, and where $g^{\rm TBA}_\alpha$ is 
the function which results from the equations
of \cite{LMSS}, analytically continued from the
massless regime.
In section \ref{sec:tbaseries} we further conjectured that 
\[
  g^{\rm CFT}_\alpha(l)
= g^{\rm TBA}_\alpha(l)
+ \fract 14 L(0)
- \fract 12 \epsilon(\fract{i\pi}2)
+ \tilde k(l)
\;,
\]
where $\tilde k(l)$ is an (unknown) regular function of $l^{12/5}$
vanishing at $l=\infty$, and which is independent of the boundary 
conditions,
and where again $g^{\rm TBA}_\alpha$ is analytically continued from the
massless regime.
This is just the sort of relation that would arise if the TBA
prediction of \cite{LMSS} has neglected some corrections arising as
next-to-leading order terms in the saddle-point evaluation of the path
integral giving the partition function. As yet we have not been able
to check this possibility.

One of the unresolved problems we encountered was the correct contour
prescription when the kernel function $\phi_\alpha$ contained
parameter-dependent poles. The result we found in section
\ref{sec:ambiguity} is that it is
consistent to take the contour to encircle those poles which
correspond to boundary bound states.
As mentioned there, it will be interesting to arrive
at this result from first principles and check it in further models.
However it is clear that no change in this contour prescription can
correct the TBA \gfs\ of \cite{LMSS} which will still have the wrong
net overall change in $g$ along the massive flows.

We should also like to point out that the \gfs\ we obtain to not
satisfy the `$g$--theorem', but since this model is non-unitary that
is perhaps no surprise. 

Another point to mention is that, in the massive case, $Y(\theta,r)$ possesses
square-root singularities in the parameter $r$.
These singularities lead  to branch-points connecting the
bulk vacuum energy to excited states \cite{DTa}, for which the
Y--functions are described by so-called excited state TBA equations
\cite{BLZ2,DTa}. 
It would be nice to understand the role of these  ``excited'' branches
of $Y$ in the framework of boundary theories, especially in the light
of the result \ref{eq:I2c}
{}From the discussion in \cite{BLZ2} it seems clear that in the {\em
massless} case the excited branches $Y_n$ give the coefficients of the
boundary state expanded in the eigenstates of the conformal field
theory conserved quantities, 
i.e. if $\cev {\psi_n}$ is the $n$-th excited state of the conformal
field theory and $\vec{\Phi(h)}$ is the boundary state for the
massless perturbation of the boundary $\Phi$, then 
(cf. eq.  (\ref{eq:gmassless}))
\[
  \veev{\psi_0}{\Phi(h)} 
= \veev{\varphi}{\Phi(h)} 
= Y(\fract{\pi\, \hat{b}}{6}) \veev{\varphi}{\One}
\;,\;\;\;\;
  \veev{\psi_n}{\Phi(h)} 
= Y_n(\fract{\pi\, \hat{b}}{6}) \veev{\psi_n}{\One}
\;.
\]
Quite how this generalises to the massive model is not yet clear,
but
for the ground state we have given strong evidence for 
\[
  \frac{\veev{\Omega}{\Phi(h)} }{\veev{\Omega}{\One}}
= Y(i\pi\fract{b+3}6) 
\]
so it is reasonable to suppose that, by analytic continuation in
$\lambda$, the same should remain true in general, i.e.
\[
  \frac{\veev{\psi_n}{\Phi(h)} }{\veev{\psi_n}{\One}}
= Y_n(i\pi\fract{b+3}6)
\;.
\]
Some preliminary numerical and CPT checks support this conjecture.

The TBA for \gfs\ was used in a recent paper by
Lesage et al.~\cite{LSS}, as part of an analysis of 
boundary flows in minimal models. In the light of the doubts raised
above, it is possible that some of their conclusions for cases where
the bulk as well as the boundary
is perturbed should now be re-examined. However, since we have
deliberately focussed on just one model in this paper, we will leave
these questions for future work.

Al.~Zamolodchikov has recently found the exact correspondence
between the UV and IR parameters in the Sine-Gordon model based on
calculations of the extensive term $f_\alpha$ from the TBA \gfs\ and
from exact results for the expectation value of the boundary field
\cite{ZamoBol}. 
Since the Yang--Lee model can be viewed as one special case of the
Sine-Gordon model, his results should include our result
relating $h$, $M$ and $b$, summarised in equations
(\ref{eq:hnum}) and (\ref{eq:hcexact}), as a special case.
So far, his results are not directly comparable with ours,
but they are of the same functional form, only $h_c$ not being
identified. 
A complete comparison between his results and ours would require the
relation between the normalisations of the fields in the two
approaches, and this has yet to be 
found.

\bigskip
\bigskip
\noindent{\bf Acknowledgements --- }
The work was supported in part by a TMR grant of the European
Commission, contract reference ERBFMRXCT960012, in part by a NATO
grant, number CRG950751, and in part by an EPSRC grant GR/K30667. PED
and GMTW thank the EPSRC for Advanced Fellowships, and RT thanks 
the Universiteit van Amsterdam for a post-doctoral
fellowship.
IR thanks the DAAD, EPSRC and KCL for
financial support.

GMTW would like to thank J.-B.~Zuber for numerous discussions;
PED and GMTW would like to thank Al.~Zamolodchikov for discussions of
these results and those of \cite{ZamoBol}.

\bigskip

\small
\renewcommand\baselinestretch{0.95}

%
%

\begin{thebibliography}{99}
%
\raggedright
\parskip 1pt
%
\bibitem{Card4}
J.L.\ Cardy,
{\it Boundary conditions, fusion rules and the Verlinde formula},
\newblock
Nucl.~Phys.~{\bf B324} (1989) 581--596.

\bibitem{AL91}
I.\ Affleck and A.W.W.\ Ludwig,
{\it
Universal noninteger ``Ground-State Degeneracy'' in critical quantum
systems},
\PRL{67} (1991) 161--164.

\bibitem{LMSS}
A. LeClair, G. Mussardo, H. Saleur and S. Skorik,
\newblock
{\it Boundary energy and boundary states in integrable quantum field
theories},
\newblock  Nucl.\ Phys.\ {\bf B453} (1995) 581--618\xtra{9503227}

\bibitem{Us1}
P.\ Dorey, A.\ Pocklington, R.\ Tateo and G.\ Watts,
{\it
TBA and TCSA with boundaries and excited states},
\NP{B525} (1998) 641--663\xtra{9712197}

\bibitem{BPPZ}
R.E. Behrend, P.A. Pearce, V.B. Petkova and J.-B. Zuber,
{\it 
Boundary Conditions in Rational Conformal Field Theories,}
\newblock {\tt hep-th/9908036.}

\bibitem{RW}
I.\ Runkel and G.\ Watts,
in preparation.

\bibitem{CLew1}
J.L. Cardy and D.C.\ Lewellen,
{\it Bulk and boundary operators in conformal field theory},
\newblock
Phys.\ Lett.\ {\bf B259} (1991) 274--278.

\bibitem{Lewe1}
D.C.\ Lewellen,
{\it Sewing constraints for conformal field theories on surfaces with
boundaries},
\newblock Nucl.\ Phys.\ {\bf B372} (1992) 654--682.

\bibitem{Runk1}
I. Runkel,
\newblock
{\it 
Boundary structure constants for the A-series Virasoro minimal
models},
\newblock \NP{B549} (1999) 563-578\xtra{9811178}

\bibitem{CPes1}
J.L. Cardy and I. Peschel,
{\it
Finite size dependence of the free energy in two-dimensional critical
systems},
\NP{B300} (1988) 377--392

\bibitem{YZam1}
V.P.\ Yurov and \AlBZ,
{\it
Truncated conformal space approach to the scaling Lee-Yang model,
}
\IJMP{A5} (1990) 3221-3246.

\bibitem{Zb}
\AlBZ, 
{\it
Thermodynamic Bethe Ansatz in Relativistic Models. Scaling
 3-state Potts and Lee-Yang Models}, 
\NP{B342} (1990) 695-720.

\bibitem{Zg}
\AlBZ, Mass scale in sine-Gordon model and its reductions,
\IJMP{A10} (1995) 1125-1150.

\bibitem{BLZ1}
V.V.\ Bazhanov, S.L.\ Lukyanov and  A.B.\ Zamolodchikov,
{\it
Integrable Structure of Conformal Field Theory, Quantum KdV Theory and
Thermodynamic Bethe Ansatz},
\CMP{177} (1996) 381--398\xtra{9412229}

\bibitem{FS1}
P. Fendley and H. Saleur, 
\newblock
{\it Deriving boundary S matrices},
\newblock  Nucl.\ Phys.\ {\bf B428} (1994) 681--693\xtra{9402045}

\bibitem{ZZ}
\ABZ\ and \AlBZ,
{\it 
Factorized S--matrices in two dimensions as the exact solutions of
certain relativistic quantum field theory models,}
{Ann. Phys.} {\bf 120} (1979) 253--291.

\bibitem{CM}
J.L. Cardy and G.Mussardo,
\newblock
{\it S matrix of the Yang-Lee edge singularity in two dimensions},
\newblock  Phys.\ Lett.\ {\bf B225} (1989) 275--278.

\bibitem{Ga}
S.\ Ghoshal,
Boundary state boundary S-matrix of the sine-Gordon model,
\IJMP{A9} (1994) 4801--4810\xtra{9310188}

\bibitem{GZ}
S. Ghoshal and A.B. Zamolodchikov,
\newblock
{\it Boundary S matrix and boundary state in two-dimensional
integrable quantum field theory},
\newblock  Int.\ J.\ Mod.\ Phys.\ {\bf A9}  (1994) 3841-3886\xtra{9306002}

\bibitem{DTa}
P.\ Dorey and R.\ Tateo,
Excited states by analytic continuation of TBA equations,
\NP{B482} (1996) 639--659\xxtra{9607167};\\
{}~~~---~~
Excited states in some simple perturbed conformal field theories,
\NP{B489} (1998) 575-623\xtra{9706140}

\bibitem{Us1b}
P.~Dorey, R.~Tateo and G.M.T.~Watts,
{\em
Generalisations of the Coleman-Thun mechanism and boundary reflection
factors},
\newblock \PL{B448} (1999) 249-256\xtra{9810098}

\bibitem{BLZ2}
V.V.\ Bazhanov, S.L.\ Lukyanov and A.B.\ Zamolodchikov,
Integrable quantum field theories in finite volume: excited state
energies,
\NP{B489} (1997) 487--531\xtra{9607099}

\bibitem{LSS}
F. Lesage, H. Saleur and P. Simonetti,
\newblock
{\it  Boundary flows in minimal models},
\newblock Phys.\ Lett.\ {\bf B427} (1998) 85--92\xtra{9802061}

\bibitem{ZamoBol}
Al.B.~Zamolodchikov,
talk at the 4th Bologna workshop on conformal and integrable models, 
June 30 - July 3, 1999 [unpublished].
%
\end{thebibliography}
\end{document}